\documentclass[journal, onecolumn,12pt, draftclsnofoot]{IEEEtran}

\usepackage{mathrsfs}

\usepackage[noadjust]{cite}
\usepackage{graphicx,color,overpic,psfrag}
\usepackage{amsmath, amssymb}
\usepackage{latexsym}
\usepackage{bm}
\usepackage{amssymb}
\usepackage{cases}
\usepackage{array}
\usepackage{fancyhdr}
\usepackage{setspace}

\ifCLASSOPTIONcompsoc
\usepackage[caption=false,font=normalsize,labelfont=sf,textfont=sf]{subfig}
\else
\usepackage[caption=false,font=footnotesize]{subfig}
\fi

\usepackage{url}
\usepackage{algpseudocode}
\usepackage{algorithm}

\usepackage{multirow}
\usepackage{dsfont}
\usepackage{tabularx}
\usepackage[table]{xcolor}

\usepackage{amsfonts}

\usepackage{letltxmacro}

\graphicspath{{figure/}}

\newtheorem{remark}{Remark}

\newtheorem{prop}{Proposition}

\newcommand{\figref}[1]{Fig. \ref{#1}}

\newcommand{\alref}[1]{Algorithm \ref{#1}}
\newcommand{\appref}[1]{Appendix \ref{#1}}
\newcommand{\secref}[1]{Section \ref{#1}}

\newcommand{\propref}[1]{Proposition \ref{#1}}

\newcommand{\Exp}{{\mathbb{E}}}
\newcommand{\expect}[1]{\Exp\left\{#1\right\}}

\newcommand{\diag}[1]{\mathsf{diag}\left\{#1\right\}}

\newcommand{\delfunc}[1]{\delta\left(#1\right)}

\newcommand{\thetabs}[2]{{\dnnot{\theta}{bs}}}

\newcommand{\equaa}{\mathop{=}^{(\mathrm{a})}}

\newcommand{\cF}{\mathcal{F}}

\newcommand{\bG}{\mathbf{G}}
\newcommand{\bH}{\mathbf{H}}
\newcommand{\bI}{\mathbf{I}}

\newcommand{\bK}{\mathbf{K}}

\newcommand{\bQ}{\mathbf{Q}}

\newcommand{\bV}{\mathbf{V}}

\newcommand{\bX}{\mathbf{X}}

\newcommand{\bzero}{\mathbf{0}}

\newcommand{\bLambda}{{\boldsymbol\Lambda}}

\newcommand{\bomega}{{\boldsymbol\omega}}

\newcommand{\dnnot}[2]{#1_{\mathrm{#2}}}

\newcommand{\ntb}{\notag\\}

\newcommand{\figsincolwid}{12cm}

\newcommand{\E}{\mathbb{E}}
\newcommand{\R}{\mathbb{R}}
\newcommand{\C}{\mathbb{C}}
\newcommand{\Q}{\mathbf{Q}}

\newcommand{\I}{\mathbf{I}}
\newcommand{\K}{\cal{K}}

\newcommand{\Lda}{\mathbf{\Lambda}}
\newcommand{\Hk}{\mathbf{H}_{k}}

\newcommand{\GkH}{\mathbf{G}_{k}^{H}}
\newcommand{\Gk}{\mathbf{G}_{k}}

\newcommand{\Qk}{\mathbf{Q}_{k}}

\newcommand{\lambdak}{\mathbf{\Lambda}_{k}}

\newcommand{\trr}{\mathrm{tr}}
\newcommand{\EE}{\mathrm{EE}}
\newcommand{\Pc}{P_{\mathrm{c}}}
\newcommand{\Ps}{P_{\mathrm{s}}}

\newcommand{\Pmax}{P_{\mathrm{max}}}
\newcommand{\kbar}{{{\overline{\mathbf{K}}}_{k}}}
\newcommand{\kt}{{{\widetilde{\mathbf{K}}}_{k}}}
\newcommand{\Rbar}{{\overline{R}}}

\allowdisplaybreaks

\begin{document}

\title{Energy Efficiency Optimization for Downlink Massive MIMO With Statistical CSIT}

\author{
Li~You, Jiayuan~Xiong,
Xinping~Yi, Jue~Wang, Wenjin~Wang, and~Xiqi~Gao
\thanks{This work was presented in part at the 2019 IEEE Global Communications Conference (GLOBECOM), Waikoloa, HI, USA.
}
\thanks{
L. You, J. Xiong, W. Wang, and X. Q. Gao are with the National Mobile Communications Research Laboratory, Southeast University, Nanjing 210096, China, and also with the Purple
Mountain Laboratories, Nanjing 211100, China (e-mail: liyou@seu.edu.cn; jyxiong@seu.edu.cn; wangwj@seu.edu.cn; xqgao@seu.edu.cn).
}
\thanks{
X. Yi is with the Department of Electrical Engineering and Electronics, University of Liverpool, Liverpool L69 3BX, U.K. (e-mail: xinping.yi@liverpool.ac.uk).
}
\thanks{
J. Wang is with the School of Information Science and Technology, Nantong University, Nantong 226019, China, also with the Research Center of Networks and Communications, Peng Cheng Laboratory, Shenzheng 518066, China, and also with the Nantong Research Institute for Advanced Communication Technologies (NRIACT), Nantong 226019, China (e-mail: wangjue@ntu.edu.cn).
}
}

\maketitle

\begin{abstract}
We investigate energy efficiency (EE) optimization for single-cell massive multiple-input multiple-output (MIMO) downlink transmission with only statistical channel state information (CSI) available at the base station. We first show that beam domain transmission is favorable for energy efficiency in the massive MIMO downlink, by deriving a closed-form solution for the eigenvectors of the optimal transmit covariance matrix. With this conclusion, the EE optimization problem is reduced to a real-valued power allocation problem, which is much easier to tackle than the original large-dimensional complex matrix-valued precoding design problem. We further propose an iterative water-filling-structured beam domain power allocation algorithm with low complexity and guaranteed convergence, exploiting the techniques from sequential optimization, fractional optimization, and random matrix theory. Numerical results demonstrate the near-optimal performance of our proposed statistical CSI aided EE optimization approach.
\end{abstract}

\newpage
\section{Introduction}
Recent years have witnessed the dramatically increasing demands of wireless data services for sheer number of mobile devices due to the emerging applications in virtual reality, cloud-based services, artificial intelligence, etc. Such data-hungry applications require higher data transmission rate for massive connections and therefore pose new challenges in future wireless communications. Thanks to the deployment of large-scale antenna array at the base stations (BSs), massive multiple-input multiple-output (MIMO) could serve a large number of user terminals (UTs) with the same time/frequency resources \cite{Marzetta10Noncooperative}.
Owing to the potential significant performance gain in spectral efficiency, massive MIMO has received tremendous attention and is deemed a disruptive and promising technology for next-generation wireless communications \cite{Sun2015Beam,marzetta2016fundamentals,bjornson2016massive,bjornson2017massive}.

Energy-aware optimization for wireless communications has received extensive research interest in the last few years, owing to the increase in both ecological and economic concerns \cite{zappone2015energy,bjornson2016deploying,wang2016energy,Zappone2016energy,pizzo2018network,Vaezy2019energy}. Conventionally, spectral efficiency was deemed to be a more important design objective than energy efficiency (EE) as data rate was a major concern given the limited radio spectrum.
However, with the ever rapid growing number of connected UTs, the power consumption could be significantly increased and EE becomes an inevitable consideration. Compared to the spectral efficiency oriented wireless transmission design, one typical performance criterion of the EE oriented approaches aims to maximize the ratio of the achievable rate to the corresponding power consumption.

Extensive works have been emerging for energy efficient wireless transmission design in traditional small scale MIMO systems \cite{xu2013energy,he2014coordinated,he2015energy,zappone2015energy,tervo2017energy,tervo2018energy}. It is worth mentioning that most existing works rely on the knowledge of instantaneous channel state information at the transmitter (CSIT). In practice, acquiring instantaneous CSIT is usually challenging in the massive MIMO downlink. For instance, relying on channel reciprocity, downlink CSI acquisition can be done via uplink training in time-division duplex (TDD) systems. However, the obtained downlink CSI may still be inaccurate due to practical limitations such as the calibration error in the radio frequency chains \cite{choi2014downlink}. Even worse, for frequency-division duplex (FDD) systems, the acquisition of the downlink CSI becomes more challenging without channel reciprocity \cite{wang2016energy}. The feedback overhead for CSI acquisition increases linearly with the number of transmit antennas when orthogonal pilot sequences are adopted, which might be unaffordable for practical massive MIMO systems \cite{You15Pilot,You16Channel}. Moreover, when the UTs are moving fast, the acquired CSI easily becomes outdated if the feedback delay is larger than the channel coherence time. Compared with instantaneous CSI, the statistical CSI, e.g., the spatial correlation and channel mean, is more likely to be stable for a longer period. Therefore, when instantaneous CSIT is not available, statistical CSI can be exploited for precoder design.
Note that it is in general not a difficult task for the BS to obtain the relatively slowly-varying statistical CSI through long-term feedback or covariance extrapolation \cite{wang2013precoder,wang2012statistical,khalilsarai2019fdd}.

To this end, we investigate energy efficient massive MIMO downlink transmission assuming that only the statistical CSIT is available. The main contributions of this paper are summarized as follows\footnote{Part of the contributions has been organized as a conference paper and submitted to IEEE Globecom 2019 \cite{xiong2019energy}.}
\begin{itemize}
\item We investigate the optimal transmission strategy for EE maximization in massive MIMO downlink and derive a necessary condition that the optimal transmit covariance matrices must follow. We show that as the number of transmit antennas grows to infinity, the eigenvectors of the optimal energy efficiency transmit covariance matrix asymptotically become unique, which are irrelevant to particular channel realizations of the UTs. As a consequence, beam domain transmission becomes favorable for statistical CSI aided energy efficient massive MIMO downlink transmission.
\item Guided by the above insight, we propose a power allocation algorithm to maximize the system EE for massive MIMO downlink in the beam domain. Exploiting the minorization-maximization (MM) algorithm, the EE maximization problem with a non-convex fractional objective is transformed into a series of concave-convex programs. Then we solve the concave-convex maximization problem by reformulating it as a series of concave programs through fractional programming. To further reduce the computational complexity, we derive a deterministic equivalent (DE) of the system EE to simplify the computation in the proposed algorithm.
\item To handle the transformed concave-convex maximization problem in a computationally efficient and well-structured way, we utilize the alternating optimization approach and decompose it into two subproblems: an EE maximization problem without the power constraint and a sum-rate maximization problem with the power constraint. For both subproblems, we propose efficient iterative water-filling-structured algorithms with guaranteed convergence. Numerical results illustrate the effectiveness of our proposed EE maximization iterative algorithm.
\end{itemize}

The rest of the paper is organized as follows. In \secref{sec:system model}, we introduce the system model. In \secref{sec:opt_EE_design}, we investigate the optimal transmission strategy design for EE optimization with statistical CSIT only. We show that the beam domain transmission is favorable for EE optimization and an energy efficient algorithm is proposed for massive MIMO downlink. In \secref{sec:low-complexity}, we further develop a low-complexity and well-structured algorithm for the EE optimization power allocation problem. The simulation results are drawn in \secref{sec:numerical_results}. The conclusion is presented in \secref{sec:conclusion}.

We adopt the following notations throughout the paper. Upper-case bold-face letters are matrices and lower-case bold-face letters are column vectors, respectively. We use ${\mathbf{I}}_M$ to denote the $M \times M$ identity matrix where the subscript is omitted when no confusion caused. The superscripts ${( \cdot )^{ - 1}}$, ${( \cdot )^T}$, and ${( \cdot )^H}$ represent the matrix inverse, transpose, and conjugate-transpose operations, respectively. The ensemble expectation, matrix trace, and determinant operations are represented by $\E \left\{  \cdot  \right\}$, $\mathrm{tr} (\cdot )$, and $\mathrm{det} ( \cdot )$, respectively. The operator $\diag{\bf{x}}$ indicates a diagonal matrix with $\bf{x}$ along its main diagonal. We use ${[{\bf{A}}]_{m,n}}$ to represent the $(m,n)$th element of matrix ${\bf{A}}$. The inequality ${\bf{A}} \succeq \bzero$ means that ${\bf{A}}$ is Hermitian positive semi-definite. The notation $[x]^+$ denotes $\max(x, 0)$. The operator $\odot$ denotes the Hadamard product. The notation $\triangleq$ is used for definitions.

\section{System Model}\label{sec:system model}
Consider a single-cell massive MIMO downlink where one BS with $M$ antennas simultaneously transmits to $K$ UTs with $N_k$ receive antennas at each UT $k \in {\cal K} \triangleq \left\{ {1,2, \ldots ,K} \right\}$.
The transmitted signal is denoted by ${\bf{x}} \in {\C^{M \times 1}}$. Note that ${\bf{x}} = \sum\nolimits_{k} {{{\bf{x}}_k}}$, where ${\bf{x}}_k$ is the signal for the $k$th UT which satisfies $ \E \{ {\bf{x}}\}  = \bzero$ and $ \E \{ {{\bf{x}}_k}{\bf{x}}_{k'}^H\}  = {\bf{0}}\left( {k' \ne k} \right)$. The transmit covariance matrix for the $k$th UT is denoted by ${{\bf{Q}}_k} = \E \{ {{\bf{x}}_k}{\bf{x}}_k^H\}  \in {\C ^{M \times M}}$. The signal received at UT $k$ is given by
\begin{align}
{{\bf{y}}_k} = {{\bf{H}}_k}{\bf{x}} + {{\bf{n}}_k}\in {\C ^{N_k \times 1}}
\end{align}
where ${\bf{H}}_k\in {\C ^{N_k \times M}} $ represents the downlink channel matrix from the BS to UT $k$ and ${{\mathbf{n}}_{k}}\in {{\mathbb{C}}^{N_k\times 1}}$ denotes the circularly symmetric complex Gaussian noise with zero mean and covariance ${\sigma ^2}{{\mathbf{I}}_{N_k}}$.

Consider the MIMO channel model with jointly correlated Rayleigh fading \cite{Gao09Statistical}, the downlink channel matrix $\Hk$ can be written as
\begin{align}\label{eq:beam_H}
{{\bf{H}}_k} = {{\bf{U}}_k}{{\bf{G}}_k}{\bf{V}}_k^H
\end{align}
where ${{\bf{U}}_k} \in {\C ^{N_k \times N_k}}$ and ${{\bf{V}}_k} \in {\C ^{M\times M}}$ are both deterministic unitary matrices, representing the eigenvectors of the receive correlation matrix and the BS correlation matrix of $\bH_{k}$, respectively. Note that ${{\bf{G}}_k} \in {\C ^{N_k \times M}}$ in \eqref{eq:beam_H} is referred in the literature as the beam domain channel matrix \cite{You17BDMA}, whose elements are zero-mean and independently distributed. The statistical CSI of ${{\bf{G}}_k}$, i.e., the eigenmode channel coupling matrix \cite{Gao09Statistical}, is modeled as
\begin{align}
{{\bf{\Omega }}_k} = \E \left\{ {{\bf{G}}_k} \odot {\bf{G}}_k^* \right\}  \in { \R ^{N_k \times M}}.
\end{align}

For massive MIMO channels, as $M \to \infty $, ${{\bf{H}}_k}$ in \eqref{eq:beam_H} can be well approximated as \cite{Adhikary13Joint,You15Pilot,You16Channel}
\begin{align}\label{eq:Hk}
{{\bf{H}}_k}\mathop  = \limits^{M \to \infty } {{\bf{U}}_k}{{\bf{G}}_k}{{\bf{V}}^H}.
\end{align}
It has been shown in \cite{You15Pilot} that ${\bf{V}}$ is independent of the locations of UTs and only depends on the BS antenna array geometry. For example, with the uniform linear arrays (ULAs) with antenna spacing of half-wavelength, the discrete Fourier transform (DFT) matrix can be used to well approximate ${\bf{V}}$ \cite{You15Pilot,You16Channel}.

It is supposed that each UT $k$ has access to instantaneous CSI of its own channel with properly designed pilot signals \cite{Sun2015Beam}, along with the covariance matrix ${{\bf{K}}_k}$ for ${{\mathbf{n}}_k'}{\rm{ = }}\sum\nolimits_{i \ne k} {{{\bf{H}}_k}{{\bf{x}}_i}}  + {{\bf{n}}_k}$, i.e., the aggregate interference-plus-noise. At each UT, we treat ${{\mathbf{n}}_k'}$ as Gaussian noise for a worst-case design \cite{Hassibi2003How} with covariance
\begin{align}
{{\bf{K}}_k} = {\sigma ^2} {\bf{I}}_{N_k} + \sum\limits_{i \ne k}^K {\E \{ {{\bf{H}}_k}{{\bf{Q}}_i}{{\bf{H}}_k^H}\} } \in { \C ^{N_k \times N_k}}.
\end{align}
Then, an ergodic data rate of the $k$th UT is given by \cite{Lu2019Robust,wu18beam}
\begin{align}\label{eq:ergodic_rate}
{R_k} & = \E \{ \log \det ({{\bf{K}}_k} + {{\bf{H}}_k}{{\bf{Q}}_k}{\bf{H}}_k^H)\}  -  \log\det({{\bf{K}}_k})\ntb
& \equaa  \E \{ \log\det({\kt} + {{\bf{G}}_k}{{\bf{V}}^H}{{\bf{Q}}_k}{\bf{VG}}_k^H)\}  -  \log\det({\kt})
\end{align}
where (a) follows from first rewriting ${\bf{H}}_k$ in terms of \eqref{eq:Hk}, then applying Sylvester's determinant identity, i.e., $\det ({\bf{I}} + {\bf{XY}}) = \det ({\bf{I}} + {\bf{YX}})$. Moreover, $\kt$ in \eqref{eq:ergodic_rate} is defined as
\begin{align}\label{eq:K_tilde}
{\kt} &\triangleq {\bf{U}}_k^H{{\bf{K}}_k}{{\bf{U}}_k}\ntb
& = {\sigma ^2}{\bf{I}}_{N_k} + \sum\limits_{i \ne k}^K {\underbrace {\E \{ {{\bf{G}}_k}{{\bf{V}}^H}{{\bf{Q}}_i}{\bf{VG}}_k^H\} }_{ \triangleq {{\bf{\Pi}}_k}({{\bf{V}}^H}{{\bf{Q}}_i}{\bf{V}})}}  \in { \C ^{N_k \times N_k}}.
\end{align}
Note that ${{\bf{\Pi}}_k}({\bf{X}})$ defined in \eqref{eq:K_tilde} is a matrix-valued function of $\bX$ with the elements given by
\begin{align}\label{eq:A}
{\left[ {{{\bf{\Pi}}_k}({\bf{X}})} \right]_{s,t}}
& = {\left[ \E\left\{\Gk{\mathbf{X}}{\mathbf{G}}_k^H\right\} \right]}_{s,t} \ntb
& = \E \left\{  {\left[ \Gk \right]}_{s,:} {\mathbf{X}} {\left[ {\mathbf{G}}_k^H \right]}_{:,t}   \right\}\ntb
& = \trr \left( \E \left\{ {\left({\left[ {\mathbf{G}}_k \right]}_{t,:}\right)}^H  {\left[ \Gk \right]}_{s,:}\right\} {\mathbf{X}}  \right)\ntb
& \equaa \trr \left( \diag{\left( {\left[{\bf{\Omega}}_k \right]}_{s,:}\right)^T}  {\mathbf{X}} \right)\cdot\delfunc{s-t}
\end{align}
where (a) follows from the fact that the elements of the beam domain channel matrix ${\mathbf{G}}_k$ are independently distributed.

We consider an affine power consumption model \cite{xu2013energy}. In particular, the total power consumed is comprised of three parts, i.e.,
\begin{align}\label{eq:power_consumption_model}
P_{\mathrm{tot}} = \xi \sum\limits_{k=1}^K{ {\mathop{\rm tr}\nolimits} (\Qk)} + M{\Pc} +\Ps
\end{align}
where the scaling coefficient $\xi$ describes the transmit amplifier inefficiency, $\sum\nolimits_{k}{ {\mathop{\rm tr}\nolimits} (\Qk)}$ represents the total transmit power, $\Pc$ denotes the dynamic power dissipations per antenna (e.g., power consumption in the digital-to-analog converter, the frequency synthesizer, the BS filter and mixer), which is independent of $\sum\nolimits_{k}{ {\mathop{\rm tr}\nolimits} (\Qk)}$, and $\Ps$ incorporates the static circuit power consumption, which is independent of both $M$ and $\sum\nolimits_{k}{ {\mathop{\rm tr}\nolimits} (\Qk)}$. In practice, $\Pc$ and $\Ps$ are usually much larger compared to the power dissipations for processing transmit signals, which can be omitted in model \eqref{eq:power_consumption_model}.

In the following, we investigate the precoding strategy design for massive MIMO downlink transmission under the EE maximization criterion, which is formulated as
\begin{align}\label{eq:EE_problem_Q}
\mathcal{F}:\quad\underset{{{\Q}_{1}},{{\Q}_{2}},\ldots ,{{\Q}_{K}}}{\mathop{\max }}\,&\quad \EE \triangleq \frac{\sum_{k=1}^{K} {{R_k}}  }{P_{\mathrm{tot}}} \ntb
{\mathrm{s.t.}}\quad
&\quad \sum\limits_{k=1}^{K}{\trr\left( \Qk \right)}\le P_{\mathrm{\max }}\ntb
&\quad \Qk \succeq \bzero,\quad\forall k\in {\cal K}
\end{align}
where the system EE is defined under the above modeling of the ergodic rate in \eqref{eq:ergodic_rate} and the power consumption in \eqref{eq:power_consumption_model}, and $P_{\mathrm{\max }}$ is constrained by the BS power budget.

\section{EE Optimal Transmission Design}\label{sec:opt_EE_design}
\begin{figure}
\centering
\includegraphics[width=0.9\textwidth]{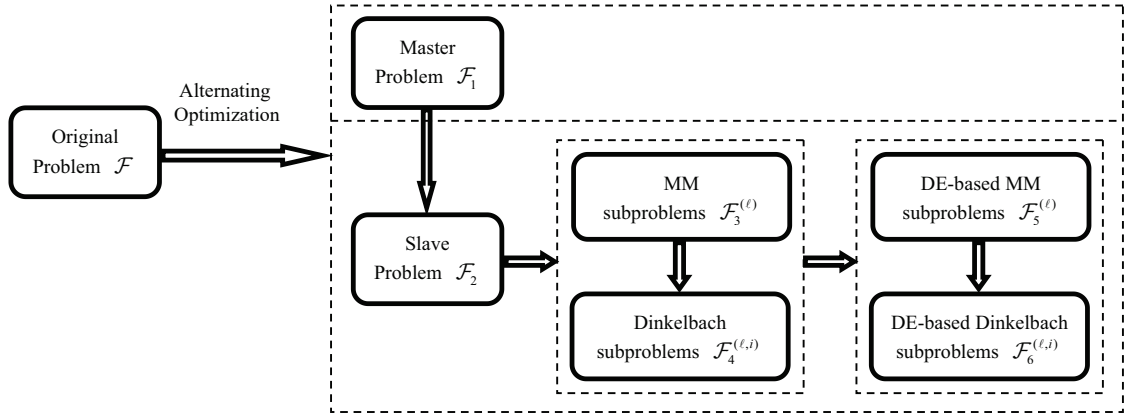}
\caption{Illustration of the solution methodology based on the alternating optimization approach.}
\label{fig:methodology}
\end{figure}

In this section, we study the optimal transmit strategy for the EE maximization problem $\mathcal{F}$ in \eqref{eq:EE_problem_Q}.
Directly solving the problem is computationally challenging. Alternatively, we can handle it more efficiently via alternating optimization by exploiting the inherent structure of problem $\mathcal{F}$. Specifically, the original problem $\mathcal{F}$ can be decomposed into smaller-sized subproblems, i.e., a master problem and a slave problem, which can be handled more easily with much fewer numbers of variables. Furthermore, these problems can be transformed into more manageable subproblems by optimization techniques. The overall solution methodology for the EE maximization problem $\mathcal{F}$ is summarized in \figref{fig:methodology}, which will be described in details in the following.

To figure out the optimal transmit strategy design for $\mathcal{F}$, we first decompose the transmit covariance matrix into ${{\bf{Q}}_k} = {{\bf{\Psi }}_k}{{\bf{\Lambda }}_k}{\bf{\Psi }}_k^H$ by eigenvalue decomposition. Note that the columns of ${\bf{\Psi }}_k$ and the corresponding diagonal elements in ${\bf{\Lambda }}_k$ are the eigenvectors and the eigenvalues of $\bQ_{k}$, respectively. In fact, the eigenmatrix (i.e., the matrix consisting of all eigenvectors) ${\bf{\Psi }}_k$ represents the subspace where the transmit signals lie in. Moreover, the elements of the diagonal matrix ${\bf{\Lambda }}_k$ represent the power assigned to each dimension/direction of the subspace for the transmit signals.
By doing so, we obtain an equivalent formulation of \eqref{eq:EE_problem_Q} as follows
\begin{align}\label{eq:decomposition}
\mathcal{F}:\quad\underset{\left\{ {{\bf{\Psi }}_k}, {\bf{\Lambda }}_k \right\}}{\mathop{\max }}&\quad \EE \ntb
{\mathrm{s.t.}}
& \quad {{\bf{\Psi }}_k} {{\bf{\Psi }}_k^H} = \I_M  \ntb
&\quad \sum\limits_{k=1}^{K}{\trr\left( \Lda_k \right)}\le P_{\mathrm{\max }}, \quad \Lda_k \succeq \bzero,\quad \forall k\in {\cal K}.
\end{align}

To obtain the optimal ${{\bf{Q}}_k}$, we could decompose $\mathcal{F}$ into a master problem $\mathcal{F}_1$ with respect to ${\bf{\Psi }}_k$ and a slave problem $\mathcal{F}_2$ with respect to ${\bf{\Lambda }}_k$, while for each problem, the other variables are considered to be fixed.
In the following, the optimization of the eigenmatrix ${{\bf{\Psi }}_k}$ and the power allocation matrix ${\bf{\Lambda }}_k$, for all UT $k$, will be respectively investigated by performing alternating optimization between the master and the slave problems.

\subsection{Optimal Transmission Direction}
In the master problem, we aim to figure out the optimal ${{\bf{\Psi }}_k}$ given the knowledge of ${\bf{\Lambda }}_k$, i.e.,
\begin{align}\label{eq:master}
\mathcal{F}_1:\quad\underset{\left\{ {{\bf{\Psi }}_k}\right\}}{\mathop{ \max }}&\quad \EE \ntb
{\mathrm{s.t.}}
& \quad {{\bf{\Psi }}_k} {{\bf{\Psi }}_k^H} = \I_M , \quad \forall k\in {\cal K}.
\end{align}
Taking advantage of the massive MIMO channel characteristics, we could identify the optimal eigenmatrix ${{\bf{\Psi }}_k}$ of the transmit signal covariance ${{\bf{Q }}_k}(\forall k)$ in the proposition as follows.
\begin{prop}\label{theorem:beam_domain_optimal}
For any power allocation matrix ${\bf{\Lambda }}_k$, the optimal eigenmatrix ${{\bf{\Psi }}_k}$ for problem $\mathcal{F}_1$ is given by the eigenmatrix ${\bf{V}}$ under channel model \eqref{eq:Hk}, i.e.,
\begin{align}
{\bf{Q}}_k^{\mathrm{opt}} = {\bf{V}}{{\bf{\Lambda }}_k}{{\bf{V}}^H},\quad \forall k \in {\cal K}.
\end{align}
\end{prop}

\begin{IEEEproof}
See \appref{app:A}.
\end{IEEEproof}

\propref{theorem:beam_domain_optimal} reveals that, to maximize the system EE in $\mathcal{F}$, the optimal directions for the downlink transmit signals should be aligned with the eigenvectors of the BS correlation matrices. In other words, the beam domain transmission is favorable for EE optimization in downlink massive MIMO.

By \propref{theorem:beam_domain_optimal}, it turns out that the optimal solution to the master problem $\mathcal{F}_1$ does not depend on ${\bf{\Lambda }}_k$, meaning that the iteration between the master and the slave problem is not necessary, hence we can solve the master problems once for all. By first optimizing ${{\bf{\Psi }}_k}$ according to \propref{theorem:beam_domain_optimal}, we can therefore characterize the slave problem as
\newcounter{mytempeqncnt}
\begin{align}\label{eq:EE_problem_lambda}
\mathcal{F}_2:\quad\underset{\Lda}{\mathop{\max }}\,\quad&\frac{ \sum\limits_{k=1}^{K}{\left(  \underbrace{\E\left\{ \log \det \left( \kbar\left(\Lda\right)+\Gk\lambdak\GkH \right) \right\}}_{\triangleq R_{k}^{+}\left( \Lda \right)}- \underbrace{\log \det \left( \kbar\left(\Lda\right) \right)}_{\triangleq R_{k}^{-}\left( \Lda \right)} \right)}}{\xi \sum\limits_{k=1}^{K}{\trr\left( \lambdak \right)} + M \Pc + \Ps} \ntb
{\mathrm{s.t.}}\quad
& \sum\limits_{k=1}^{K}{\trr\left( \lambdak \right)}\le P_{\mathrm{\max }} \ntb
& \lambdak \succeq \bzero,\quad \forall k\in \K
\end{align}
where $\Lda \triangleq \left\{ {{\Lda}_{1}},{{\Lda}_{2}},\ldots ,{{\Lda}_{K}} \right\}$ and
\begin{align}
\kbar\left(\Lda\right) \triangleq {\sigma ^2} \I_{N_k} +  \sum\limits_{i\ne k}^{K}{{{\bf{\Pi}}_k}({{\bf{\Lambda }}_i})}.
\end{align}

With this manipulation, the slave problem $\mathcal{F}_2$ now turns to be a power allocation problem in the beam domain. Note that the number of optimization variables is reduced from $M^{2}K$ in original problem $\mathcal{F}$ to $MK$ in slave problem $\mathcal{F}_2$.
Therefore, $\mathcal{F}_2$ is much simpler compared with the original matrix-valued energy efficient precoding design $\mathcal{F}$. In the following, we proceed to investigate efficient power allocation to solve the slave problem $\mathcal{F}_2$.

\subsection{Optimal Power Allocation}

To tackle problem $\mathcal{F}_2$ where the objective is fractional, we adopt a fractional programming method. Note that $R_{k}^{+}\left( \Lda \right)$ and $R_{k}^{-}\left( \Lda \right)$ defined in $\mathcal{F}_2$ are both concave over $\Lda$, leading to a non-concave numerator of the objective in $\mathcal{F}_2$. Consequently, directly utilizing classical fractional programming approaches would exhibit an exponential complexity \cite{zappone2015energy}. In fact, $\mathcal{F}_2$ is NP-hard in the sense that there do not exist algorithms with polynomial-time complexity to guarantee the globally optimal solution.

In the following, we develop an efficient approach to derive the EE maximization power allocation strategies by means of sequential convex optimization tools with fractional programming methods \cite{Marks1978A,Beck2010A,Razaviyayn2012A}. More specifically, we resort to the MM algorithm \cite{sun2017majorization} to handle $\mathcal{F}_2$ and the main idea of MM algorithm lies in converting a non-convex problem to a series of easy-to-handle subproblems. From $\mathcal{F}_2$, we can find that the numerator of the objective is the difference between two concave functions.
Denoting by $\triangle R_{k,\mathrm{ub}}^-$ the first-order Taylor expansion of the negative rate term $R_{k}^{-}\left( \Lda \right)$, we have $R_k^-(\Lda) \le \triangle R_{k,\mathrm{ub}}^-$.
Then replace the negative rate term $R_{k}^{-}\left( \Lda \right)$ with its first-order Taylor expansion, the numerator in each slave problem can be lower-bounded by a concave function.
By doing so, the slave problem $\mathcal{F}_2$ is tackled through solving a series of fractional subproblems as follows
\begin{align}\label{eq:CCCP_ergodic}
\mathcal{F}_3^{(\ell)}:\quad \underset{\Lda}{\mathop{\max }}\,\quad&\frac{ \sum\limits_{k=1}^{K}{\left(  R_{k}^{+}\left(\Lda\right)- R_{k}^{-}\left(\Lda^{(\ell)}\right) - {{\mathop{\rm tr}\nolimits} \left( {{\left( {\frac{\partial }{{\partial {{\bf{\Lambda }}_k}}}\sum\limits_{k'=1}^K {R_{k'}^{-}\left(\Lda^{(\ell)}\right) } } \right)}^{T} \left( {{{\bf{\Lambda }}_k} - {{\bf{\Lambda }}^{(\ell)}_{k}}} \right)} \right)} \right)} }{ \xi \sum\limits_{k=1}^{K}{\trr\left( \lambdak \right)} + M \Pc + \Ps }\ntb
{\mathrm{s.t.}}\quad
& \sum\limits_{k=1}^{K}{\trr\left( \lambdak \right)}\le P_{\mathrm{\max }}\ntb
& \lambdak \succeq \bzero,\quad\forall k\in \K
\end{align}
where $\Lda^{\left( \ell \right)} \triangleq \left\{ {{\Lda}^{(\ell)}_1},{{\Lda}^{(\ell)}_2},\ldots ,{{\Lda}^{(\ell)}_K} \right\}$ with $\ell$ denoting the iteration index. For the objective of the subproblem $\mathcal{F}_3^{(\ell)}$, the linearization of $R_k^ - ({\bf{\Lambda }})$ by the first-order Taylor expression makes the numerator of the objective function concave over ${\bf{\Lambda }}$. By doing so, we now focus on the concave-convex fractional optimization problem $\mathcal{F}_3^{(\ell)}$ and tackle it in polynomial complexity by directly using fractional programming theories \cite{zappone2015energy}.

In this paper, Dinkelbach's algorithm is used to solve $\mathcal{F}_3^{(\ell)}$, owing to its advantage of not having additional constraints compared with Charnes-Cooper algorithm \cite{shen2018fractional}. Specifically, $\mathcal{F}_3^{(\ell)}$ is equivalently solved by a series of concave subproblems in the following
\begin{align}\label{eq:Dinkelbach_ergodic}
\mathcal{F}_4^{(\ell,i)}:\quad \Lda^{(\ell)}_{\left(i+1\right)} = \underset{\Lda}{\mathop{\arg\max }}\, \quad &{ \sum\limits_{k=1}^{K}{\left(  R_{k}^{+}\left(\Lda\right) - R_{k}^{-}\left(\Lda^{(\ell)}\right) - {{\mathop{\rm tr}\nolimits} \left( {{{{\bf{\Delta }}^{(\ell)}_{k}}} \left( {{{\bf{\Lambda }}_k} - {{\bf{\Lambda }}^{(\ell)}_{k}}} \right)} \right)} \right)} } \ntb
  & \qquad - {\eta}^{(\ell)}_{(i)}\left({ \xi \sum\limits_{k=1}^{K}{\trr\left( \lambdak \right)} + M \Pc + \Ps }\right) \ntb
{\mathrm{s.t.}} \quad
& \sum\limits_{k=1}^{K}{\trr\left( \lambdak \right)}\le P_{\mathrm{\max }} \ntb
& \lambdak \succeq \bzero,\quad\forall k\in \K
\end{align}
where $\Lda^{\left(i+1\right)}_{\left( \ell \right)} \triangleq \left\{ {{\Lda}^{\left(i+1\right)}_{1,\left(\ell\right)}},{{\Lda}^{\left(i+1\right)}_{2,\left(\ell\right)}},\ldots ,{{\Lda}^{\left(i+1\right)}_{K,\left(\ell\right)}} \right\}$ and ${{\bf{\Delta }}^{(\ell)}_k}$ denotes the derivative of $\sum\nolimits_{k'} {R_{k'}^{-}\left(\Lda^{(\ell)}\right) }$ over ${{{\bf{\Lambda }}_k}}$ as
\begin{align}\label{eq:derivative_CCCP}
{{\bf{\Delta }}^{(\ell)}_{k}}  = \sum\limits_{k' \ne k} {\sum\limits_{n = 1}^{N_{k'}} {\frac{{{{\widehat {\bf R}}_{k',n}}}}{{ {\sigma}^2 + {\mathop{\rm tr}\nolimits} ({{\bf{\Lambda }}^{(\ell)}_{{\backslash k'}}}{{\widehat {\bf R}}_{k',n}})}}} }
\end{align}
where ${{{\bf{\Lambda }}^{(\ell)}_{{\backslash k'}}}}= \sum\nolimits_{i \ne k'} {{{\bf{\Lambda }}^{(\ell)}_{i}}}$ and ${{\widehat {\bf R}}_{k',n}} = \diag{ {\bomega } _{k',n}}$ with ${{\bomega } _{k',n}^{T}}$ being the $n$th row of ${{\bf{\Omega }}_{k'}}$. Note that ${{\bf{\Delta }}^{(\ell)}_{k}}$ is a diagonal matrix with the corresponding $t$th diagonal entry given by
\begin{align}
{[{{\bf{\Delta }}^{(\ell)}_{k}}]_{t,t}} = \sum\limits_{k' \ne k} {\sum\limits_{n = 1}^{N_{k'}} {\frac{{{{[{{\bf{\Omega }}_{k'}}]}_{n,t}}}}{{{\sigma}^2 +  \sum\limits_{i \ne k'}^K {\sum\limits_{m = 1}^M {{{[{{\bf{\Lambda }}^{(\ell)}_{i}}]}_{m,m}}{{[{{\bf{\Omega }}_{k'}}]}_{n,m}}} } }}} }.
\end{align}
The auxiliary variable ${\eta}^{(\ell)}_{(i)}$ in \eqref{eq:Dinkelbach_ergodic} can be iteratively updated as
\begin{align}\label{eq:eta_ergodic}
{\eta}^{(\ell)}_{(i)} = \frac{ \sum\limits_{k=1}^{K}{\left(  R_{k}^{+}\left( \Lda^{(\ell)}_{(i)} \right) - R_{k}^{-}\left(\Lda^{(\ell)}\right) - {{\mathop{\rm tr}\nolimits} \left( {{{{\bf{\Delta }}^{(\ell)}_{k}}} \left( {{{\bf{\Lambda }}^{(\ell)}_{k,(i)}} - {{\bf{\Lambda }}^{(\ell)}_{k}}} \right)} \right)} \right)}}{ \xi \sum\limits_{k=1}^{K}{\trr\left( {{\bf{\Lambda }}_{k,(i)}^{(\ell)}} \right)} + M \Pc + \Ps }
\end{align}
with $i$ denoting the iteration index. Note that the parametric problem in \eqref{eq:Dinkelbach_ergodic} required to be addressed in each iteration is concave, thereby, it can be handled by applying classical convex optimization approaches \cite{Boyd04Convex}.
Moreover, it can be readily proved that the Dinkelbach-based method can converge to the globally optimal solution of the fractional problem $\mathcal{F}_3^{(\ell)}$ \cite{shen2018fractional}.

We can conclude that the sequence of the objective values generated by $\mathcal{F}_3^{(\ell)}$ converges, which follows from the convergence properties of MM method \cite{sun2017majorization}. Moreover, every limit point of the sequence is a local optimum of problem $\mathcal{F}_2$.
However, when calculating the numerator in $\mathcal{F}_3^{(\ell)}$ in each iteration, i.e., the system sum-rate, manipulating the expectation operation through Monte-Carlo methods is quite computationally cumbersome. To avoid this, we use the random matrix theory \cite{Couillet11Random,Lu16Free} to replace the rate expression by its deterministic equivalent. More specifically, the DE of $R_{k}^{+}\left( \Lda \right)$ is computed by
\begin{align}\label{eq:DE}
\Rbar_{k}^{+}\left( \Lda \right) = \log \det \left( \mathbf{I}_{M}+\mathbf{\Gamma }_{k}{{\mathbf{\Lambda }}_{k}} \right) +\log \det \left( \mathbf{\widetilde{\Gamma }}_{k}+\mathbf{\overline{K}}_{k}\left( \Lda \right) \right) -\operatorname{tr}\left( \mathbf{I}_{N_k}-{{\mathbf{\widetilde{\Phi }}}^{-1}_{k}} \right)
\end{align}
where
\begin{align}
\label{eq:gamma_1}
\mathbf{\Gamma }_{k}&={{\bf{\Xi}}_{k}}\left( {{{\mathbf{\widetilde{\Phi }}}^{-1}_{k}}}{ {\left(\mathbf{\overline{K}}_{k}\left( \Lda \right)\right)} ^{-1}} \right)\\
\label{eq:gamma_2}
\mathbf{\widetilde{\Gamma }}_{k}&={{\bf{\Pi}}_{k}}\left( { {{{\mathbf{\Phi }}^{-1}_{k}}}}{{ {\mathbf{\Lambda }}_{k}} } \right)\\
\label{eq:phi_1}
\mathbf{\widetilde{\Phi }}_{k}&=\mathbf{I}_{N_k}+{{\bf{\Pi}}_{k}}\left( { {{{\mathbf{\Phi }^{-1}_{k}}}}}{{\mathbf{\Lambda }}_{k}} \right){{\left(\mathbf{\overline{K}}_{k}\left( \Lda \right)\right)}^{-1}}\\
\label{eq:phi_2}
\mathbf{\Phi }_{k}&=\mathbf{I}_{M}+{{\bf{\Xi }}_{k}}\left( { {{{\mathbf{\widetilde{\Phi }}}^{-1}_{k}}}}{ {\left(\mathbf{\overline{K}}_{k}\left( \Lda \right)\right)} ^{-1}} \right){{\mathbf{\Lambda }}_{k}}\\
\label{eq:Xi}
{{\bf{\Xi}}_k} & \left( {\bf{X}} \right) \triangleq \E \left\{ {\bf{G}}_k^H {\bf{X}} \Gk \right\}.
\end{align}
Similarly to the procedure of deriving the elements of ${{\bf{\Pi}}_{k}}({\bf{X}})$ in \eqref{eq:A}, the elements of ${{\bf{\Xi}}_k}({\bf{X}})$ defined in \eqref{eq:Xi} can be written as
\begin{align}\label{eq:Xi_element}
{\left[ {{{\bf{\Xi}}_k}({\bf{X}})} \right]_{s,t}}
= \trr \left( \diag{{\left[{\bf{\Omega}}_k \right]}_{:,s}} {\mathbf{X}} \right)\cdot\delfunc{s-t}.
\end{align}
From \eqref{eq:DE} to \eqref{eq:phi_2}, we observe that the DE expression $\Rbar_{k}^{+}\left( \Lda \right)$ depends mainly on ${{\bf{\Xi}}_{k}}({\bf{X}})$ and ${{\bf{\Pi}}_{k}}(\widetilde{{\bf{X}}})$, which can be both efficiently computed.
Moreover, the concavity of $\Rbar_{k}^{+}\left( \Lda \right)$ over $\Lda$ can be concluded from \cite{Dumont2010On,Dupuy2011On}. Then, with the aid of \eqref{eq:DE}, we turn to the following concave-convex fractional subproblems
\begin{align}\label{eq:CCCP_DE}
\mathcal{F}_5^{(\ell)}:\quad\underset{\Lda}{\mathop{\max }}\,\quad &\frac{ \sum\limits_{k=1}^{K}{\left(  \Rbar_{k}^{+}\left( \Lda \right)- R_{k}^{-}\left( \Lda^{(\ell)} \right) - {{\mathop{\rm tr}\nolimits} \left( {{{\bf{\Delta }}^{(\ell)}_{k}} \left( {{{\bf{\Lambda }}_k} - {{\bf{\Lambda }}^{(\ell)}_{k}}} \right)} \right)} \right)} }{ \xi \sum\limits_{k=1}^{K}{\trr\left( \lambdak \right)} + M \Pc + \Ps }\ntb
{\mathrm{s.t.}}\quad
& \sum\limits_{k=1}^{K}{\trr\left( \lambdak \right)}\le P_{\mathrm{\max }} \ntb
& \lambdak \succeq \bzero, \quad \forall k\in \K.
\end{align}
Note that the solution to $\mathcal{F}_5^{(\ell)}$ is an asymptotically optimal solution to $\mathcal{F}_3^{(\ell)}$. Utilizing Dinkelbach's transform, $\mathcal{F}_5^{(\ell)}$ can be solved by considering a series of concave subproblems as follows
\begin{align}\label{eq:Dinkelbach_DE}
\mathcal{F}_6^{(\ell,i)}:\quad \underset{\Lda}{\mathop{\max }}\,\quad&{ \sum\limits_{k=1}^{K}{\left(  \Rbar_{k}^{+}\left(\Lda\right) - R_{k}^{-}\left( \Lda^{(\ell)} \right) - {{\mathop{\rm tr}\nolimits} \left( {{{{\bf{\Delta }}^{(\ell)}_{k}}} \left( {{{\bf{\Lambda }}_k} - {{\bf{\Lambda }}^{(\ell)}_{k}}} \right)} \right)} \right)} } \ntb
 & \qquad - {\overline \eta}^{(\ell)}_{(i)}\left( { \xi \sum\limits_{k=1}^{K}{\trr\left( \lambdak \right)} + M \Pc + \Ps } \right)\ntb
{\mathrm{s.t.}}\quad
& \sum\limits_{k=1}^{K}{\trr\left( \lambdak \right)}\le P_{\mathrm{\max }}\ntb
& \lambdak \succeq \bzero,\quad\forall k\in \K
\end{align}
where
\begin{align}\label{eq:eta_DE}
{\overline \eta}^{(\ell)}_{(i)} = \frac{ \sum\limits_{k=1}^{K}{\left(  \Rbar_{k}^{+}\left( \Lda^{(\ell)}_{(i)} \right) - R_{k}^{-}\left( \Lda^{(\ell)} \right) - {{\mathop{\rm tr}\nolimits} \left( {{{{\bf{\Delta }}^{(\ell)}_{k}}} \left( {{{\bf{\Lambda }}_{k,(i)}^{(\ell)}} - {{\bf{\Lambda }}^{(\ell)}_{k}}} \right)} \right)} \right)}}{ \xi \sum\limits_{k=1}^{K}{\trr\left( \Lda_{k,(i)}^{(\ell)} \right)} + M \Pc + \Ps }.
\end{align}

In summary, our proposed statistical CSI-aided transmission design, which jointly utilizes the MM method, the Dinkelbach's algorithm, and the DE theory, is detailed in \alref{alg:MM_DE}, where
\begin{subequations}
\begin{align}\label{eq:EEl_DE}
{\EE^{(\ell)}} & =    \frac{{\sum\limits_{k = 1}^K \left({ \Rbar_k^ + ({{\bf{\Lambda }}^{(\ell)}}) - R_k^ - ({{\bf{\Lambda }}^{(\ell)}})}\right) }}{ \xi \sum\limits_{k=1}^{K}{\trr\left( \Lda^{(\ell)}_{k} \right)} + M \Pc + \Ps } \\ \label{eq:F_DE}
F^{(\ell)}_{(i)} & =   \sum\limits_{k=1}^{K}{\left(  \Rbar_{k}^{+}\left( \Lda^{(\ell)}_{(i)} \right) - R_{k}^{-}\left( \Lda^{(\ell)} \right) - {{\mathop{\rm tr}\nolimits} \left( {{{{\bf{\Delta }}^{(\ell)}_{k}}} \left( {{{\bf{\Lambda }}_{k,(i)}^{(\ell)}} - {{\bf{\Lambda }}^{(\ell)}_{k}}} \right)} \right)} \right)}\ntb
 & \qquad- {\overline \eta}^{(\ell)}_{(i)}\left({ \xi \sum\limits_{k=1}^{K}{\trr\left( \Lda_{k,(i)}^{(\ell)} \right)} + M \Pc + \Ps }\right).
\end{align}
\end{subequations}

\begin{remark}
If only considering the numerator (i.e., by setting $\xi = 0$, $\Pc = 0$ and $\Ps = 1$), $\mathcal{F}$ becomes a non-fractional program of sum-rate maximization, and \alref{alg:MM_DE} can be utilized to perform sum-rate maximization for massive MIMO downlink transmission.
\end{remark}

\begin{algorithm}[h]\label{alg:MM_DE}
\caption{DE-Based EE Maximization Algorithm}
\label{alg:MM_DE}
\begin{algorithmic}[1]
\State Initialize $\Lda^{(0)}$, threshold ${\varepsilon}_1$, ${\varepsilon}_2$, set iteration $\ell = 0$, and calculate $\EE^{(0)}$ as \eqref{eq:EEl_DE}.
\Repeat
\State Calculate DE auxiliary matrices ${\bf{ \Gamma }}^{(\ell)}_{k}$ and ${\bf{\widetilde \Gamma }}^{(\ell)}_{k}$ for all $k$ by \alref{alg:DE}.
\State Calculate the derivative ${{\bf{\Delta }}^{(\ell)}_{k}}$ for all $k$ as \eqref{eq:derivative_CCCP}.
\State Initialize $\Lda^{(\ell)}_{(0)}$, $ {\overline \eta}^{(\ell)}_{(0)} = 0$, $ F^{(\ell)}_{(0)} >{\varepsilon}_1$, and set iteration $i = 0$.
\While{$ F^{(\ell)}_{(i)}>{\varepsilon}_1 $}
\State Solve $\mathcal{F}_6^{(\ell,i)}$ and set ${\bf{\Lambda }}^{(\ell)}_{(i+1)}$ as the solution.
\State Update ${\overline \eta}^{(\ell)}_{(i+1)}$ as \eqref{eq:eta_DE}.
\State Update $F^{(\ell)}_{(i+1)}$ as \eqref{eq:F_DE}.
\State Set $i=i+1$.
\EndWhile
\State Update $\Lda^{(\ell+1)} = \Lda^{(\ell)}_{(i)}$.
\State Set $\ell=\ell+1$, and calculate $\EE^{(\ell)}$ as \eqref{eq:EEl_DE}.
\Until{$\left|  \EE^{(\ell )} - \EE^{( \ell-1 )}  \right|\le \varepsilon_2$}
\end{algorithmic}
\end{algorithm}

\begin{algorithm}[!t]
\caption{Deterministic Equivalent Method}
\label{alg:DE}
\begin{algorithmic}[1]
\Require
Initial power allocation matrices $\Lda_1^{(\ell)},\Lda_2^{(\ell)},\ldots ,\Lda_K^{(\ell)}$ and the preset threshold ${\varepsilon}_3$.
\Ensure
DE auxiliary matrices ${\bf{ \Gamma }}^{(\ell)}_{k}$ and ${\bf{\widetilde \Gamma }}^{(\ell)}_{k}$, $k = 1,2, \ldots ,K$.
\State Initialization: ${\bf{\widetilde \Phi }}_k^{(u)}$, $u = 0$.
\Repeat
\State Calculate ${\bf{\widetilde \Phi }}_k^{(u+1)}$ and ${\bf{\Phi }}_k^{(u+1)}$ by \eqref{eq:phi_1} and \eqref{eq:phi_2}.
\State Set $u=u+1$.
\Until{$\left|  {\bf{\widetilde \Phi }}_k^{(u)} - {\bf{\widetilde \Phi }}_k^{(u-1)}  \right|\le \varepsilon_3 $}
\State Calculate ${\bf{ \Gamma }}^{(\ell)}_{k}$ and ${\bf{\widetilde \Gamma }}^{(\ell)}_{k}$ by \eqref{eq:gamma_1} and \eqref{eq:gamma_2}, $k = 1,2, \ldots ,K$.
\end{algorithmic}
\end{algorithm}

\section{Low-Complexity Power Allocation Algorithm}\label{sec:low-complexity}

The parametric problem $\mathcal{F}_6^{(\ell,i)}$ to be solved in each iteration is a concave program, which can be tackled through classical convex optimization approaches \cite{Boyd04Convex}. However, the computational complexity of the numerical methods will be high when the number of BS antennas becomes large. This calls for the development of a low-complexity method for the beam domain power allocation problem. In the following, a more efficient and well-structured iterative algorithm for $\mathcal{F}_5^{(\ell)}$ is developed.

Unlike the sum-rate maximization problem, transmission with all power budget might not be optimal for the EE maximization design, owing to the fact that the system EE will saturate when the excessive power is consumed.
Therefore, seeking the optimal transmit power consumption is critical to EE optimization. To figure out the relationship between the system EE and the transmit power, we first introduce an auxiliary function given by
\begin{align}\label{eq:f_P}
f^{(\ell+1)}(P_{\rm{T}}) \triangleq \underset{\Lda}{\mathop{\max }}\,\quad & { \sum\limits_{k=1}^{K}{\left(  \Rbar_{k}^{+}\left( \Lda \right) - R_{k}^{-}\left( \Lda^{(\ell)} \right) - {{\mathop{\rm tr}\nolimits} \left( {{{{\bf{\Delta }}^{(\ell)}_{k}}} \left( {{{\bf{\Lambda }}_k} - {{\bf{\Lambda }}^{(\ell)}_{k}}} \right)} \right)} \right)} }\ntb
{\mathrm{s.t.}} \quad
& \sum\limits_{k=1}^{K}{\trr\left( \lambdak \right)} = P_{\rm{T}} \ntb
& \lambdak \succeq \bzero,\quad\forall k\in \K
\end{align}
where $P_{\rm{T}}$ is an auxiliary power variable. Note that given an overall transmit power $P_{\rm{T}}$, $f^{(\ell+1)}(P_{\rm{T}})$ is the corresponding maximum system sum-rate. Then, we consider the following problem
\begin{align}\label{eq:varsigma}
\underset{P_{\rm{T}}}{\mathop{\arg\max }}\,\quad&{ \varsigma^{(\ell+1)} (P_{\rm{T}}) = \frac{{f^{(\ell+1)}(P_{\rm{T}})}}{\xi P_{\rm{T}} + M \Pc +\Ps} } \ntb
{\mathrm{s.t.}}\quad
& 0 \le P_{\rm{T}} \le \Pmax
\end{align}
where $f^{(\ell+1)}(P_{\rm{T}})$ is the introduced auxiliary function in \eqref{eq:f_P}. Denoting by $P_{\rm{T}}^{*}$ the optimal solution of problem \eqref{eq:varsigma}, we can then obtain that $f^{(\ell+1)}(P_{\rm{T}}^{*})$ is indeed the optimal objective value of $\mathcal{F}_5^{(\ell)}$.
Since the objective function in \eqref{eq:f_P} is concave, $f^{(\ell+1)}(P_{\rm{T}})$ is nondecreasing and concave with respect to $P_{\rm{T}}$ \cite[Lemma 5]{chong2011analytical}. In addition, the power consumption is an affine function of $P_{\rm{T}}$. Therefore, the objective $\varsigma^{(\ell+1)} (P_{\rm{T}})$ in \eqref{eq:varsigma} is a pseudo-concave function \cite{chong2011analytical}, and there exists a unique globally optimal point.
Thus, we can obtain that either $\varsigma^{(\ell+1)} (P_{\rm{T}})$ is nondecreasing in $[0,\Pmax]$, or there exits a point ${P_{\rm{opt}}^{(\ell+1)}}\in[0,\Pmax]$ that maximizes $\varsigma^{(\ell+1)} (P_{\rm{T}})$ such that $\varsigma^{(\ell+1)} (P_{\rm{T}})$ is monotonically nondecreasing when $P_{\rm{T}} < {P_{\rm{opt}}^{(\ell+1)}}$, and monotonically nonincreasing when $P_{\rm{T}} > {P_{\rm{opt}}^{(\ell+1)}}$ \cite{Boyd04Convex}.

Motivated by the above properties, we first consider the subproblem, which is similar to $\mathcal{F}_5^{(\ell)}$ but without the power constraint, as follows
\begin{align}\label{eq:P_1}
{\cal{P}}1^{(\ell+1)}: \{ {\bf{\Lambda }}^{(\ell+1)}_{k,\rm{opt}}\} _{k = 1}^K = \mathop {\arg \max }\limits_{{{\bf{\Lambda }}_k}:{{\bf{\Lambda }}_k} \succeq \bzero}\ \frac{{\sum\limits_{k = 1}^K {\left( {{\overline R}_k^{ +}\left( \Lda^{(\ell)} \right)  - R_{k}^{ -}\left( \Lda^{(\ell)} \right)  - {\mathop{\rm tr}\nolimits} \left( {{{\bf{\Delta }}^{(\ell)}_{k}}({{\bf{\Lambda }}_k} - {{\bf{\Lambda }}^{(\ell)}_{k}})} \right)} \right)} }}{{\xi \sum\limits_{k = 1}^K {{\mathop{\rm tr}\nolimits} ({{\bf{\Lambda }}_k})}  + M\Pc + \Ps}}.
\end{align}
If the power consumption corresponding to the optimal power allocation matrices ${P_{\rm{opt}}^{(\ell+1)}} = \sum\nolimits_{k} {{\mathop{\rm tr}\nolimits} \left(\Lda^{(\ell+1)}_{k,\rm{opt}}\right)}$ lies within the feasible power region $[0, \Pmax]$ of $\mathcal{F}_5^{(\ell)}$, the optimal solution of the unconstrained EE optimization problem in ${\cal{P}}1^{(\ell+1)}$ is indeed equal to that of the constrained EE optimization problem $\mathcal{F}_5^{(\ell)}$. On the other hand, if ${P_{\rm{opt}}^{(\ell+1)}}>\Pmax$, we can obtain that transmission with all power budget $\Pmax$ is EE optimal, and we consider the following subproblem
\begin{align}\label{eq:P_2}
{\cal{P}}2^{(\ell+1)}: \{ {\bf{\widetilde \Lambda }}^{(\ell+1)}_{k,\rm{opt}}\} _{k = 1}^K =  \mathop {\arg \max }\limits_{\scriptstyle{{\bf{\Lambda }}_k}:{{\bf{\Lambda }}_k} \succeq \bzero\hfill\atop
\sum\nolimits_{k} {{\mathop{\rm tr}\nolimits} ({{\bf{\Lambda }}_k})} =\Pmax \hfill} \frac{{\sum\limits_{k = 1}^K {\left( {{\overline R}_k^{+}\left( \Lda \right)  - R_{k}^ {-}\left( \Lda^{(\ell)} \right)  - {\mathop{\rm tr}\nolimits} \left( {{{\bf{\Delta }}^{(\ell)}_{k}}({{\bf{\Lambda }}_k} - {{\bf{\Lambda }}^{(\ell)}_{k}})} \right)} \right)} }}{{\xi \sum\limits_{k = 1}^K {{\mathop{\rm tr}\nolimits} ({{\bf{\Lambda }}_k})}  + M\Pc + \Ps}} \ntb
=  \mathop {\arg \max }\limits_{\scriptstyle{{\bf{\Lambda }}_k}:{{\bf{\Lambda }}_k} \succ \bzero\hfill\atop
\sum\nolimits_{k} {{\mathop{\rm tr}\nolimits} ({{\bf{\Lambda }}_k})} = \Pmax \hfill} \sum\limits_{k = 1}^K {\left( {{\overline R}_k^{+}\left( \Lda \right)  - R_{k}^{-}\left( \Lda^{(\ell)} \right)  - {\mathop{\rm tr}\nolimits} \left( {{{\bf{\Delta }}^{(\ell)}_{k}}({{\bf{\Lambda }}_k} - {{\bf{\Lambda }}^{(\ell)}_{k}})} \right)} \right)}.
\end{align}
Thus, we can establish the relationship between the solution for the EE maximization problem $\mathcal{F}_5^{(\ell)}$ and the solutions for problems ${\cal{P}}1^{(\ell+1)}$ and ${\cal{P}}2^{(\ell+1)}$ as
\begin{align}\label{eq:relpro}
{\bf{\Lambda }}_k^{(\ell+1)} = \left\{ \begin{array}{l}
{\bf{\Lambda }}^{(\ell+1)}_{k,\rm{opt}},\quad {P_{\rm{opt}}^{(\ell+1)}} \le \Pmax \\
{\bf{\widetilde \Lambda }}^{(\ell+1)}_{k,\rm{opt}},\quad {P_{\rm{opt}}^{(\ell+1)}} \ge \Pmax
\end{array} \right.,\quad k = 1,2, \ldots ,K.
\end{align}
Based on the relationship in \eqref{eq:relpro}, we present the description of our proposed low-complexity EE optimal power allocation algorithm in \alref{alg:simplified}. Subsequently, to obtain $\Lda_k^{(\ell+1)}(\forall k)$ in the $\ell$th iteration of \alref{alg:simplified}, we will derive low-complexity algorithms for the above two subproblems in the following subsections.

\begin{algorithm}[h]
\caption{Low-Complexity Power Allocation Algorithm}
\label{alg:simplified}
\begin{algorithmic}[1]
\State Initialize $\Lda^{(0)}$, threshold ${\varepsilon}_4$, set iteration $\ell = 0$, and calculate $\EE^{(0)}$ as \eqref{eq:EEl_DE}.
\Repeat
\State Calculate DE auxiliary matrices ${\bf{ \Gamma }}^{(\ell)}_{k}$ and ${\bf{\widetilde \Gamma }}^{(\ell)}_{k}$ for all $k$ by \alref{alg:DE}.
\State Calculate the derivative ${{\bf{\Delta }}^{(\ell)}_{k}}$ for all $k$ as \eqref{eq:derivative_CCCP}.
\State Solve problem ${\cal{P}}1^{(\ell+1)}$ by fractional programming and iterative water-filling in \alref{alg:EE iterative water-filling} and set $\left\{ {\bf{\Lambda }}^{(\ell+1)}_{k,\rm{opt}} \right\} _{k = 1}^K$ as the solution.
\State Calculate ${P_{\rm{opt}}^{(\ell+1)}} = \sum\limits_{k = 1}^K {{\mathop{\rm tr}\nolimits} \left({\bf{\Lambda }}^{(\ell+1)}_{k,\rm{opt}}\right)}$.
\If {${{P_{\rm{opt}}^{(\ell+1)}}} \le \Pmax $}
\State \State Set $\left\{ {{\bf{\Lambda }}^{(\ell+1)}_{k}}\right\} _{k = 1}^K = \left\{ {\Lda_{k,\rm{opt}}^{(\ell+1)}} \right\} _{k = 1}^K$.
\Else
\State Solve problem ${\cal{P}}2^{(\ell+1)}$ by iterative water-filling in \alref{alg:sum_rate_maximum} and set $\{ {\bf{\widetilde \Lambda }}^{(\ell+1)}_{k,\rm{opt}} \} _{k = 1}^K$ as the solution.
\State Set $\left\{ {{\bf{\Lambda }}^{(\ell+1)}_{k}}\right\} _{k = 1}^K = \left\{ {\bf{\widetilde \Lambda }}^{(\ell+1)}_{k,\rm{opt}} \right\} _{k = 1}^K$.
\EndIf
\State Set $\ell=\ell+1$, and calculate $ \EE^{(\ell)}$ as \eqref{eq:EEl_DE}.
\Until{$\left|  \EE^{(\ell )} - \EE^{( \ell-1 )}  \right|\le \varepsilon_4$}
\end{algorithmic}
\end{algorithm}

\subsection{Unconstrained EE Optimization Problem}
We first solve the unconstrained EE optimization subproblem ${\cal{P}}1^{(\ell+1)}$ in \eqref{eq:P_1}, whose objective is concave-convex fractional. Via invoking Dinkelbach's method, we can solve ${\cal{P}}1^{(\ell+1)}$ by iteratively solving the convex optimization subproblems as follows
\begin{align}\label{eq:unconstrained_problem}
\Lda^{(\ell)}_{(i+1)} = \underset{\Lda}{\mathop{\arg\max }}\,\quad&{ \sum\limits_{k=1}^{K}{\left(  {\Rbar}_{k}^{+}\left( \Lda \right) - R_{k}^{-}\left( \Lda^{(\ell)} \right) - {{\mathop{\rm tr}\nolimits} \left( {{{{\bf{\Delta }}^{(\ell)}_{k}}} \left( {{{\bf{\Lambda }}_k} - {{\bf{\Lambda }}^{(\ell)}_{k}}} \right)} \right)} \right)} } \ntb
& \qquad - {\overline \eta}^{(\ell)}_{(i)}\left({ \xi \sum\limits_{k=1}^{K}{\trr\left( \lambdak \right)} + M \Pc + \Ps }\right)\ntb
{\mathrm{s.t.}}\quad
& \lambdak \succeq \bzero,\quad\forall k\in \K.
\end{align}
To obtain the solution to \eqref{eq:unconstrained_problem} in the $i$th iteration of Dinkelbach's method, we have the following proposition.
\begin{prop}\label{prop:EE_solution}
The optimal power allocation matrices $\Lda^{(\ell)}_{(i+1)}$ to \eqref{eq:unconstrained_problem} are the solution to the following convex optimization problem
\begin{align}\label{eq:Dinkelbach_short}
{\bf{\Lambda }}^{(\ell)}_{(i + 1)} =  \mathop {\arg \max }\limits_{\bf{\Lambda }} \quad & \sum\limits_{k = 1}^K {\Bigg( {\log \det \left( {{\bf{I}}_{M} + {{\bf{\Gamma }}_k}{{\bf{\Lambda }}_k}} \right)}}\ntb
& \quad  { + \log \det \left( {{{{\bf{\widetilde \Gamma }}}_k} + {{{\bf{\overline K}}}_k \left( \Lda\right)}} \right) - {\mathop{\rm tr}\nolimits} \left( {{{\bf{\Delta }}^{(\ell)}_{k}}{{\bf{\Lambda }}_k}} \right) - \xi {{\overline{\eta}}^{(\ell)}_{(i)}}{\mathop{\rm tr}\nolimits} \left( {{{\bf{\Lambda }}_k}} \right)} \Bigg) \ntb
{\mathrm{s.t.}}\quad
& \lambdak \succeq \bzero,\quad\forall k\in \K.
\end{align}
The $m$th element ${{\bf{\lambda }}_{k,m,(i+1)}^{(\ell)}}$ of the solution ${{\bf{\Lambda }}_{k,(i+1)}^{(\ell)}}$ in \eqref{eq:Dinkelbach_short} satisfies
\begin{align}\label{eq:Solution_lambda}
\left\{ \begin{array}{l}
\frac{{\gamma _{k,m,(i+1)}^{(\ell)}}}{{1 + \gamma _{k,m,(i+1)}^{(\ell)}\lambda _{k,m,(i+1)}^{(\ell)}}}\\
\quad + \sum\limits_{k' \ne k}^K {\sum\limits_{n = 1}^{N_{k'}} {\frac{{{{\widehat r}_{k',m,n}}}}{{\widetilde \gamma _{k',n,(i+1)}^{(\ell)} + {\sigma}^2 + {\mathop{\rm tr}\nolimits} ({{\widehat {\bf{R}} }_{k',n}}{\bf{\Lambda }}_{\backslash k',(i+1)}^{(\ell)})}}} }  = {d^{(\ell)} _{k,m}} + \xi {{\overline \eta}^{(\ell)}_{(i)}},\quad\xi {{\overline \eta}^{(\ell)} _{(i)}} < \upsilon _{k,m,(i+1)}^{(\ell)} - {d ^{(\ell)}_{k,m}}\\
\lambda _{k,m,(i+1)}^{(\ell)} = 0,\quad\quad\quad\quad\quad\quad\quad\quad\quad\quad\quad\quad\quad\quad\quad\quad\quad\quad\xi {{\overline \eta}^{(\ell)} _{(i)}} \ge \upsilon _{k,m,(i+1)}^{(\ell)} - d ^{(\ell)}_{k,m}
\end{array} \right.
\end{align}
with the auxiliary variable $\upsilon _{k,m,(i+1)}^{(\ell)}$ expressed as
\begin{align}
& \upsilon _{k,m,(i+1)}^{(\ell)} = \gamma _{k,m,(i+1)}^{(\ell)} + \sum\limits_{k' \ne k}^K {\sum\limits_{n = 1}^{N_{k'}} {\frac{{{{\widehat r}_{k',m,n}}}}{{\widetilde \gamma _{k',n,(i+1)}^{(\ell)} + {\sigma}^2 + \sum\limits_{\scriptstyle(l',m')\hfill\atop
\scriptstyle \in {\cal{S}}(k,m,k')\hfill} {{{\widehat r}_{k',m',n}}\lambda _{l',m',(i+1)}^{(\ell)}} }}} }, \ntb
& \quad {{\cal{S}}_{k,m,k'}} = \left\{ {(l',m')|l' \ne k',(l',m') \ne (k,m),}  {l' \in \{ 1,2, \ldots ,K\} ,m' \in \{ 1,2, \ldots ,M\} } \right\}
\end{align}
where $\gamma _{k,m,(i+1)}^{(\ell)}$, ${{\widehat r}_{k',m,n}}$, and ${d ^{(\ell)}_{k,m}}$ are the $m$th diagonal elements of ${\bf{\Gamma }}_{k,(i+1)}^{(\ell)}$, ${{\widehat {\bf R}}_{k',n}}$, and ${{\bf{\Delta }}^{(\ell)}_{k}}$, respectively, and $\widetilde \gamma _{k',n,(i+1)}^{(\ell)}$ is the $n$th diagonal element of ${\bf{\widetilde \Gamma }}_{k',(i+1)}^{(\ell)}$.
\end{prop}

\begin{IEEEproof}
See \appref{app:B}.
\end{IEEEproof}

The solutions in \eqref{eq:Solution_lambda} indicate that the asymptotic-optimal power allocation matrices for all UTs follow the classical water-filling structure and the water level depends on the auxiliary variable ${{\overline \eta} ^{(\ell)}_{(i)}}$. Specifically, for the single-UT case with $K = 1$, the solutions can be obtained in closed-form, i.e., $\lambda_{k,m,(i+1)}^{(\ell)} = {\left[ {\frac{1}{{\xi {{\overline{\eta}} ^{(\ell)}_{(i)}}}} - \frac{1}{\gamma_{k,m,(i+1)}^{(\ell)} }} \right]^ + }$. For the case of multiple UTs, it is in general difficult to obtain the solutions in a closed-form. Thus, we propose an EE maximization iterative water-filling algorithm in \alref{alg:EE iterative water-filling}, where the auxiliary variables $\rho _{k,m,(i)}^{(\ell)}({\overline x_{k,m}})$ and ${\rho '}_{k,m,(i)}^{(\ell)}({\overline x_{k,m}})$ in Step 8 are defined as
\begin{align}\label{eq:rho_1}
\rho _{k,m,(i)}^{(\ell)}({\overline x_{k,m}})  = &\frac{{\gamma _{k,m,(i)}^{(\ell)}}}{{1 + \gamma _{k,m,(i)}^{(\ell)}{\overline x_{k,m}}}} - d^{(\ell)}_{k,m} - \xi {{\overline{\eta}}_{(i)} ^{(\ell)}} \ntb
&+\sum\limits_{k' \ne k}^K { \sum\limits_{n = 1}^{N_{k'}} {\frac{{{{\widehat r}_{k',m,n}}}}{{\widetilde \gamma _{k',n,(i)}^{(\ell)} + {\sigma}^2 +  {{{\widehat r}_{k',m,n}}{\overline x_{k,m}}} +  \sum\limits_{\scriptstyle \ \ (l',m')\hfill\atop
\scriptstyle \in {\cal{S}}(k,m,k')\hfill} {{{\widehat r}_{k',m',n}}x_{l',m'}} }}} }\\ \label{eq:rho_2}
{\rho '}_{k,m,(i)}^{(\ell)}({\overline x_{k,m}}) = & - \frac{{{{(\gamma _{k,m,(i)}^{(\ell)})}^2}}}{{{{(1 + \gamma _{k,m,(i)}^{(\ell)}{\overline x_{k,m}})}^2}}}\ntb
 &-  \sum\limits_{k' \ne k}^K {\sum\limits_{n = 1}^{N_{k'}} {\frac{{\widehat r_{k',m,n}^2}}{{{{(\widetilde \gamma _{k',n,(i)}^{(\ell)} + {\sigma}^2 +  {{\widehat r}_{k',m,n}}{\overline x_{k,m}} +  \sum\limits_{\scriptstyle \ \ (l',m')\hfill\atop
\scriptstyle \in {\cal{S}}(k,m,k')\hfill} {{{\widehat r}_{k',m',n}}x_{l',m'}} )}^2}}}} }
\end{align}
respectively. Note that \alref{alg:EE iterative water-filling} is a generalized water-filling algorithm. For our considered multi-UT scenario, accurately solving \eqref{eq:Solution_lambda} is usually challenging, which is caused by the summation of fractional functions. Therefore, Newton-Raphson method \cite{cormen2009introduction} is utilized to acquire the approximate solutions of \eqref{eq:Solution_lambda} in Step 9, i.e., the $m$th element ${{\bf{\lambda }}_{k,m,(i+1)}^{(\ell)}}$ of ${{\bf{\Lambda }}_{k,(i+1)}^{(\ell)}}$. Note that problem \eqref{eq:Dinkelbach_short} is concave and \eqref{eq:Solution_lambda} is derived from the corresponding Karush-Kuhn-Tucker (KKT) conditions. Then even when \eqref{eq:Solution_lambda} has multiple solutions, the objective values corresponding to different solutions are the same. For the single-UT case, with the explicit solutions of \eqref{eq:Solution_lambda} instead of the iterative Newton-Raphson method, \alref{alg:EE iterative water-filling} reduces to a standard water-filling algorithm.

\begin{algorithm}[h]
\caption{EE Maximum Iterative Water-Filling Algorithm}
\label{alg:EE iterative water-filling}
\begin{algorithmic}[1]
\State Initialize $\Lda_{(0)}^{(\ell)}=\Lda^{(\ell)}$, threshold $\varepsilon_5$, and $\varepsilon_6$, set iteration $i = 0$, and calculate ${\overline{\eta}}^{(\ell)}_{(0)}$ as \eqref{eq:eta_DE}.
\Repeat
\State Initialize diagonal matrices ${{\bf{X}}}_k = {{\bf{\Lambda }}_{k,(i)}^{(\ell)}}$, $k=1,2,\ldots,K $. Here, $x_{k,m}$ is the $m$th diagonal entries of ${\bf{X}}_k$.
\For{$k=1$ to $K$}
\For{$m=1$ to $M$}
\State Set ${w_{k,m}}=0$ and $\overline x_{k,m}^{(w_{k,m})}=x_{k,m}^{(w_{k,m})}$.
\Repeat
\State Calculate $\rho _{k,m,(i)}^{(\ell)}(\overline x_{k,m}^{({w_{k,m}})})$ and ${\rho '}_{k,m,(i)}^{(\ell)}(\overline x_{k,m}^{({w_{k,m}})})$ by \eqref{eq:rho_1} and \eqref{eq:rho_2}.
\State Update $\overline x_{k,m}$ as $\overline x_{k,m}^{({w_{k,m}} + 1)} = \overline x_{k,m}^{({w_{k,m}})} - \rho _{k,m,(i)}^{(\ell)}(\overline x_{k,m}^{({w_{k,m}})})/{\rho '}_{k,m,(i)}^{(\ell)}(\overline x_{k,m}^{({w_{k,m}})})$.
\State Set $w_{k,m}=w_{k,m}+1$.
\Until{$\left| {\overline x_{k,m}^{({w_{k,m}})} - \overline x_{k,m}^{({w_{k,m}} - 1)}} \right| \le {\varepsilon _5}$}
\EndFor
\EndFor
\State Update ${{x}_{k,m}} = {\left[ {\overline x_{k,m}^{({w_{k,m}})}} \right]^ + }$.
\State Update ${{\bf{\Lambda }}^{(\ell)}_{k,(i + 1)}} = {{\bf{X}}_k}$, $k=1,2,\ldots,K $.
\State Calculate ${\overline{\eta}}^{(\ell)}_{(i+1)}$ as \eqref{eq:eta_DE}.
\State Set $i=i+1$.
\Until{$\left|  {\overline{\eta}}_{(i)}^{(\ell)} - {\overline{\eta}}^{(\ell)}_{( i-1 )}  \right|\le \varepsilon_6$ }
\end{algorithmic}
\end{algorithm}

\subsection{Sum-Rate Optimization Problem}
According to the relationship in \eqref{eq:relpro}, if the solution to ${\cal{P}}1^{(\ell+1)}$ is not feasible for $\mathcal{F}_5^{(\ell)}$, we then need to solve ${\cal{P}}2^{(\ell+1)}$.
Employing a similar procedure for solving \eqref{eq:unconstrained_problem}, we have the proposition on the solution to the sum-rate maximization problem ${\cal{P}}2^{(\ell+1)}$ in the $\ell$th iteration of MM method as follows.
\begin{prop}\label{prop:Rate_solution}
The optimal power allocation matrices $\Lda^{(\ell+1)}$ to ${\cal{P}}2^{(\ell+1)}$ are the solution to the following convex optimization problem
\begin{align}\label{eq:Rate_short}
{\bf{\Lambda }}^{(\ell + 1)} = \mathop {\arg \max }\limits_{{{\bf{\Lambda }}}} \quad & \sum\limits_{k = 1}^K {\left( {\log \det \left( {{\bf{I}}_{M} + {{\bf{\Gamma }}_k}{{\bf{\Lambda }}_k}} \right)}  { + \log \det \left( {{{{\bf{\widetilde \Gamma }}}_k} + {{{\bf{\overline K}}}_k \left(\Lda\right)}} \right) - {\mathop{\rm tr}\nolimits} \left( {{{\bf{\Delta }}^{(\ell)}_{k}}{{\bf{\Lambda }}_k}} \right)} \right)}\ntb
{\mathrm{s.t.}}\quad
& \lambdak \succeq \bzero,\quad\forall k\in \K.
\end{align}
The $m$th element ${{\bf{\lambda }}^{(\ell+1)}_{k,m}}$ of ${{\bf{\Lambda }}_{k}^{(\ell + 1)}}$ in \eqref{eq:Rate_short} satisfies
\begin{align}\label{eq:Solution_Rate}
\left\{ \begin{array}{l}
\frac{{\gamma _{k,m}^{(\ell+1)}}}{{1 + \gamma _{k,m}^{(\ell+1)}\lambda _{k,m}^{(\ell + 1)}}} + \sum\limits_{k' \ne k}^K {\sum\limits_{n = 1}^{N_{k'}} {\frac{{{{\widehat r}_{k',m,n}}}}{{\widetilde \gamma _{k',n}^{(\ell+1)} + {\sigma}^2 + {\mathop{\rm tr}\nolimits} ({{\widehat {\bf{R}}}_{k',n}}{\bf{\Lambda }}_{\backslash k'}^{(\ell + 1)})}}} }  = d^{(\ell)} _{k,m} + {\mu ^{(\ell + 1)}},\quad{\mu ^{(\ell + 1)}} < \upsilon _{k,m}^{(\ell + 1)}\\
\lambda _{k,m}^{(\ell + 1)} = 0, \quad\quad\quad\quad\quad\quad\quad\quad\quad\quad\quad\quad\quad\quad\quad\quad\quad\quad\quad\quad\quad\ \ {\mu ^{(\ell + 1)}} \ge \upsilon _{k,m}^{(\ell + 1)}
\end{array} \right.
\end{align}
where the Lagrange multiplier ${\mu ^{(\ell + 1)}}$ satisfies the following KKT conditions
\begin{align}\label{eq:mu_constraint}
{\mu ^{(\ell + 1)}}\left( {{\mathop{\rm tr}\nolimits} \left( {\sum\nolimits_k {{\bf{\Lambda }}_k^{^{(\ell + 1)}}} } \right) - \Pmax} \right) & = 0 \ntb
{\mu ^{(\ell + 1)}} &\geq 0
\end{align}
and the auxiliary variable $\upsilon _{k,m}^{(\ell + 1)}$ in \eqref{eq:Solution_Rate} is given by
\begin{align}
\upsilon _{k,m}^{(\ell + 1)}  = \gamma _{k,m}^{(\ell+1)}  - d^{(\ell)} _{k,m}+ \sum\limits_{k' \ne k}^K {\sum\limits_{n = 1}^{N_{k'}} {\frac{{{{\widehat r}_{k',m,n}}}}{{\widetilde \gamma _{k',n}^{(\ell+1)} + {\sigma}^2 + \sum\limits_{\scriptstyle \ \ (l',m')\hfill\atop
\scriptstyle \in {\cal{S}}(k,m,k')\hfill} {{{\widehat r}_{k',m',n}}\lambda _{l',m'}^{(\ell + 1)}} }}} }.
\end{align}
\end{prop}
Here, we omit the proof of \propref{prop:Rate_solution} for brevity since it is similar to that of \propref{prop:EE_solution}.

Note that the solutions in \eqref{eq:Solution_Rate} also follow a similar structure to the classical water-filling solution. In particular, for the single-UT case with $K = 1$, the solutions are given in the closed-form as $\lambda _{k,m}^{(\ell + 1)} = {\left[ {{{({d ^{(\ell)}_{k,m}} + {\mu ^{(\ell + 1)}})}^{ - 1}} - {{(\gamma _{k,m}^{(\ell+1)})}^{ - 1}}} \right]^ + }$. The choice of ${\mu ^{(\ell + 1)}}$ depends on the constraints in \eqref{eq:mu_constraint}. For the multi-UT case, we propose the sum-rate maximum iterative water-filling algorithm in \alref{alg:sum_rate_maximum} to efficiently solve ${\cal{P}}2^{(\ell+1)}$, where
\begin{align}\label{eq:nu_1}
\nu _{k,m}^{(\ell)}({\overline x_{k,m}})& =  \frac{{\gamma _{k,m}^{(\ell)}}}{{1 + \gamma _{k,m}^{(\ell)}{\overline x_{k,m}}}} - d^{(\ell)} _{k,m} - \mu^{(\ell)} \ntb
&\qquad+  \sum\limits_{k' \ne k}^K { \sum\limits_{n = 1}^{N_{k'}} {\frac{{{{\widehat r}_{k',m,n}}}}{{\widetilde \gamma _{k',n}^{(\ell)} + {\sigma}^2 +  {{{\widehat r}_{k',m,n}}{\overline x_{k,m}}} +  \sum\limits_{\scriptstyle \ \ (l',m')\hfill\atop
\scriptstyle \in {\cal{S}}(k,m,k')\hfill} {{{\widehat r}_{k',m',n}}x_{l',m'}} }}} }\\ \label{eq:nu_2}
{\nu '}_{k,m}^{(\ell)}({\overline x_{k,m}})  & =  - \frac{{{{(\gamma _{k,m}^{(\ell)})}^2}}}{{{{(1 + \gamma _{k,m}^{(\ell)}{\overline x_{k,m}})}^2}}}\ntb
 &\qquad -  \sum\limits_{k' \ne k}^K {\sum\limits_{n = 1}^{N_{k'}} {\frac{{\widehat r_{k',m,n}^2}}{{{{(\widetilde \gamma _{k',n}^{(\ell)} + {\sigma}^2 +  {{\widehat r}_{k',m,n}}{\overline x_{k,m}} +  \sum\limits_{\scriptstyle \ \ (l',m')\hfill\atop
\scriptstyle \in {\cal{S}}(k,m,k')\hfill} {{{\widehat r}_{k',m',n}}x_{l',m'}} )}^2}}}} } \\ \label{eq:mu_max}
\mu_{\max} & = \mathop {\max }\limits_{k,m}\ \gamma _{k,m}^{(\ell)} +  \sum\limits_{k' \ne k}^K {\sum\limits_{n = 1}^{N_{k'}} {\frac{{{{\widehat r}_{k',m,n}}}}{{\widetilde \gamma _{k',n}^{(\ell)} + {\sigma}^2}} - d^{(\ell)} _{k,m}} }.
\end{align}
Note that \alref{alg:sum_rate_maximum} is also a generalized water-filling algorithm and Newton-Raphson method is exploited to acquire the approximate roots of \eqref{eq:Solution_Rate}. In addition, the bisection method is utilized to find the optimal ${\mu ^{(\ell + 1)}}$ under constraints \eqref{eq:mu_constraint} in \alref{alg:sum_rate_maximum}.

\begin{algorithm}[h]
\caption{Sum-Rate Maximum Iterative Water-Filling Algorithm}
\label{alg:sum_rate_maximum}
\begin{algorithmic}[1]
\State Initialize diagonal matrices ${{\bf{X}}}_k = {{\bf{\Lambda }}_k^{(\ell)}}$, $k=1,2,\ldots,K $ and threshold $\varepsilon_7$ and $\varepsilon_8$. Here, $x_{k,m}$ is the $m$th diagonal elements of ${\bf{X}}_k$. Initialize ${\mu}^{(u')}_{\min}=0$ and ${\mu}^{(u')}_{\max}$ by \eqref{eq:mu_max}, set iteration $u'=0$, and calculate ${\mu ^{(u')}}= \frac{1}{2} ({\mu _{\max}^{(u')}}+{\mu _{\min}^{(u')}})$.
\Repeat
\For{$k=1$ to $K$}
\For{$m=1$ to $M$}
\State Set ${w_{k,m}}=0$ and $\overline x_{k,m}^{(w_{k,m})}=x_{k,m}^{(w_{k,m})}$.
\Repeat
\State Calculate $\nu _{k,m}^{(\ell)}(\overline x_{k,m}^{({w_{k,m}})})$ and ${\nu  '}_{k,m}^{(\ell)}(\overline x_{k,m}^{({w_{k,m}})})$ by \eqref{eq:nu_1} and \eqref{eq:nu_2}.
\State Update $\overline x_{k,m}$ as $\overline x_{k,m}^{({w_{k,m}} + 1)} = \overline x_{k,m}^{({w_{k,m}})} - \nu _{k,m}^{(\ell)}(\overline x_{k,m}^{({w_{k,m}})})/{\nu '}_{k,m}^{(\ell)}(\overline x_{k,m}^{({w_{k,m}})})$.
\State Set $w_{k,m}=w_{k,m}+1$.
\Until{$\left| {\overline x_{k,m}^{({w_{k,m}})} - \overline x_{k,m}^{({w_{k,m}} - 1)}} \right| \le {\varepsilon _7}$}
\EndFor
\EndFor
\State Update ${{x}_{k,m}} = {\left[ {\overline x_{k,m}^{({w_{k,m}})}} \right]^ + }$ and calculate ${p_{\rm{tot}}} = \sum\limits_{k = 1}^K {\sum\limits_{m = 1}^M {{x_{k,m}}} } $.
\If{${p_{\mathrm{tot}}} < \Pmax $}
\State Set ${\mu _{\min}^{(u' + 1)}}={\mu _{\min}^{(u')}}$ and ${\mu _{\max}^{(u' + 1)}}={\mu^{(u')}}$.
\Else
\State Set ${\mu _{\min}^{(u' + 1)}}={\mu^{(u')}}$ and ${\mu _{\max}^{(u' + 1)}}={\mu _{\max}^{(u')}}$.
\EndIf
\State Update ${\mu ^{(u' + 1)}}= \frac{1}{2} ({\mu _{\max}^{(u' + 1)}}+{\mu _{\min}^{(u' + 1)}})$ and set $u'=u'+1$.
\Until{$\left| {\Pmax - {p_{\mathrm{tot}}}} \right| \le {\varepsilon _8}$}
\end{algorithmic}
\end{algorithm}

\subsection{Convergence and Complexity Analysis}
For the convergence of the proposed low-complexity algorithms, we start with the convergence of \alref{alg:EE iterative water-filling} owing to the utilization of the EE maximum iterative water-filling procedure in \alref{alg:simplified}. Firstly, during each step, since \eqref{eq:unconstrained_problem} is a concave problem, the EE maximum iterative water-filling can achieve the global maximum through solving the KKT optimality conditions \cite{Boyd04Convex}. Secondly, following from the convergence properties of Dinkelbach's method, the solution sequence $\left\{{{\bf{\Lambda }}_{1,(i)}^{(\ell)}},{{\bf{\Lambda }}_{2,(i)}^{(\ell)}}, \ldots ,{{\bf{\Lambda }}_{K,(i)}^{(\ell)}}\right\}_{i=0}^{\infty}$ converges to the global optimum \cite{zappone2015energy}. Thus, the EE maximum iterative water-filling converges to the global optimum for ${\cal{P}}1^{(\ell+1)}$. In addition, the sum-rate maximum iterative water-filling in \alref{alg:sum_rate_maximum} can converge to the global optimum for ${\cal{P}}2^{(\ell+1)}$ \cite{sun2017bdma}. Moreover, the objective value sequence ${\left\{ \EE^{(\ell)} \right\}_{\ell=0}^{\infty}}$ output by \alref{alg:simplified} is convergent based on the convergence properties of the MM method \cite{sun2017majorization}. Thus, the proposed low-complexity power allocation \alref{alg:simplified} is convergent.

Then, we discuss the complexity of our proposed algorithms. For each iteration in \alref{alg:simplified}, the optimization procedure is separated into two subproblems to obtain ${{\bf{\Lambda }}^{(\ell + 1)}}$, i.e., the unconstrained EE optimization problem ${\cal{P}}1^{(\ell+1)}$ and the sum-rate maximization problem ${\cal{P}}2^{(\ell+1)}$. Owing to the fast convergence rate of ${\bf{\widetilde \Phi }}_k^{(u+1)}$ and ${\bf{ \Phi }}_k^{(u+1)}$, and the low complexity of their calculations, the major complexity of each iteration in \alref{alg:simplified} is composed of the complexity of \alref{alg:EE iterative water-filling} or \alref{alg:sum_rate_maximum}. For \alref{alg:EE iterative water-filling}, the outer layer of ${\overline{\eta}}^{(\ell)}_{(i)}$ converges after very few iterations, which is shown by the numerical results illustrated in \secref{sec:numerical_results}. Therefore, the complexity of \alref{alg:EE iterative water-filling} depends mainly on the iterations required in the convergence of Newton-Raphson method to solve \eqref{eq:Solution_lambda}. Specifically, with a precision of $g$ digits \cite{Boyd04Convex}, the number of iterations required for Newton-Raphson method is $\log g$ \cite{cormen2009introduction}.
For \alref{alg:sum_rate_maximum}, the complexity of the inner iteration is the same as that in \alref{alg:EE iterative water-filling}. For the outer iteration, the bisection method will also converge very fast \cite{Boyd04Convex}.
Thus, the computational complexity of \alref{alg:simplified} is approximately ${\mathcal{O}}(JKM \log g + JKM)$, where $J$ is the number of iterations required for the MM method in \alref{alg:simplified}, which is usually very small as it can be observed in the simulations. Note that the complexity of \alref{alg:MM_DE} can be similarly obtained as be ${\mathcal{O}}(JLK^{3}M^{3})$ where $L$ is the number of iterations required in Dinkelbach's transform, assuming that each subproblem in \eqref{eq:Dinkelbach_DE} is solved using standard interior point methods \cite{Boyd04Convex}.
Therefore, the computational complexity of \alref{alg:simplified} can be significantly reduced when compared with \alref{alg:MM_DE}, especially for the cases with large numbers of UTs $K$ or BS antennas $M$.

\section{Numerical Results}\label{sec:numerical_results}
Numerical analysis is presented to evaluate the performance of our proposed statistical CSI aided EE optimization approach for massive MIMO downlink transmission. The QuaDRiGa channel model \cite{jaeckel2014quadriga} with a suburban macro cell scenario is adopted throughout the simulations. A total of $K=8$ UTs are randomly distributed in the cell sector. The pathloss is set as $-120$ dB for all UTs \cite{shen2018fractional}. In the simulations, the antenna array topology ULA is adopted for the BS and each UT $k$, with the number of antennas being $M=128$ and $N_k = 4$, respectively. The spacing between antennas is half-wavelength. The amplifier inefficiency factor is set as $\xi = 5$, the hardware dissipated power per antenna and the static power consumption are respectively set to $\Pc = 30$ dBm and $\Ps = 40$ dBm. The noise variance is set as ${\sigma}^2 = -105$ dBm \cite{he2013coordinated}. As the proposed algorithms converge to local optimum, the numerical results are obtained via averaging over initialization points.

The EE performance and the sum-rate performance of the approaches which aim for EE maximization and sum-rate maximization are compared in Figs. 2(a) and 2(b), respectively. The results show that, in the low power budget regime, the performance of the EE- and sum-rate-oriented approaches are almost identical, which indicates that transmission with all power budget is nearly energy efficient. In the large power budget regime, our EE-oriented approach achieved substantially better EE performance compared with the sum-rate maximization approach. This is due to the reason that there exists a threshold value of the transmit power for maximizing the system EE, thus any excess power will depress the system EE. Unlike the EE optimization design, the sum-rate optimization design always uses the overall power budget to maximize the system sum-rate regardless of the cost, which might also degrade the system EE. We can also observe that the DE results are almost identical to those obtained from the Monte-Carlo results.

\begin{figure*}[!t]
\centering
\subfloat[]{\centering\includegraphics[width=0.48\textwidth]{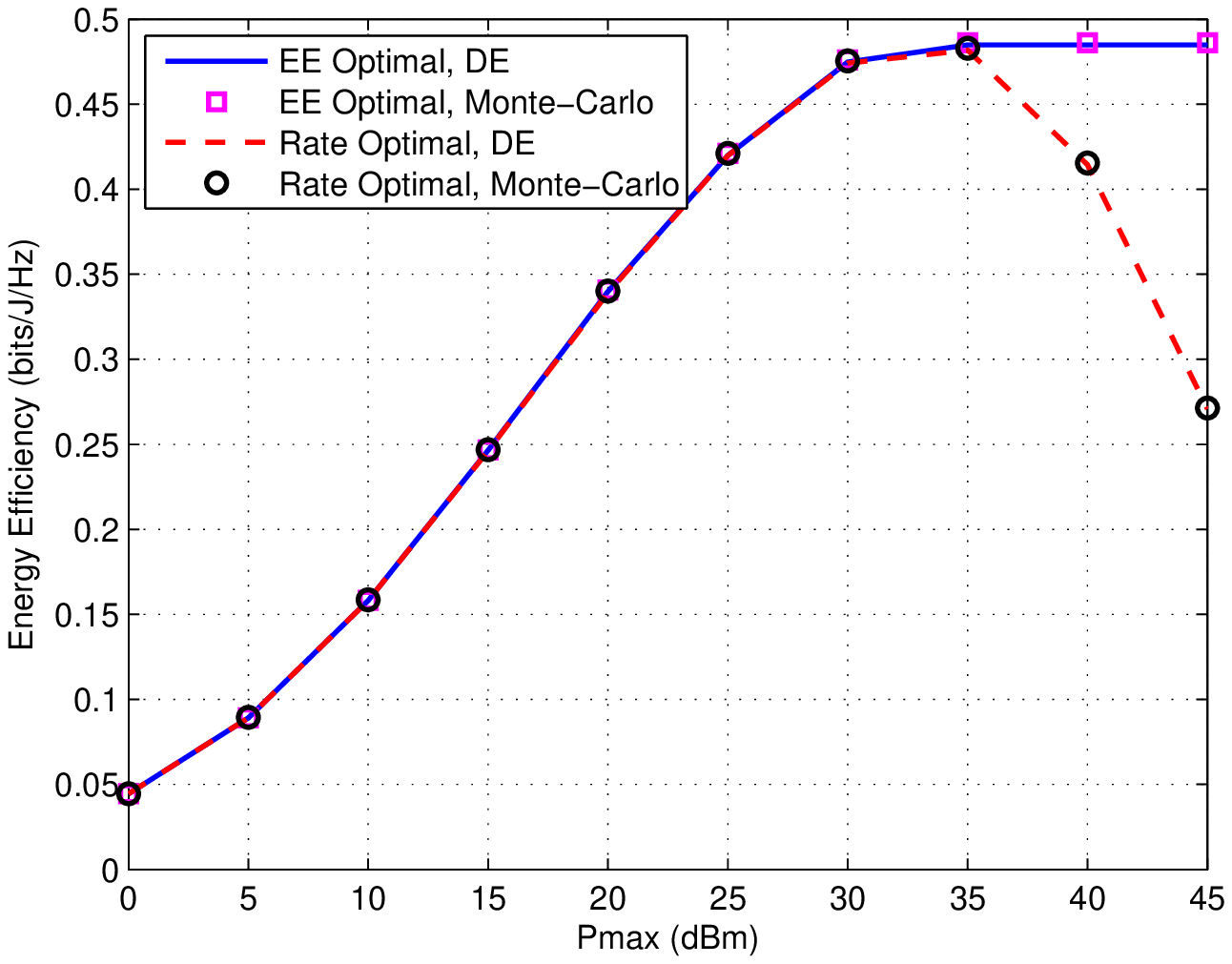}
\label{fig:EE_performance}}
\hfill
\subfloat[]{\centering\includegraphics[width=0.48\textwidth]{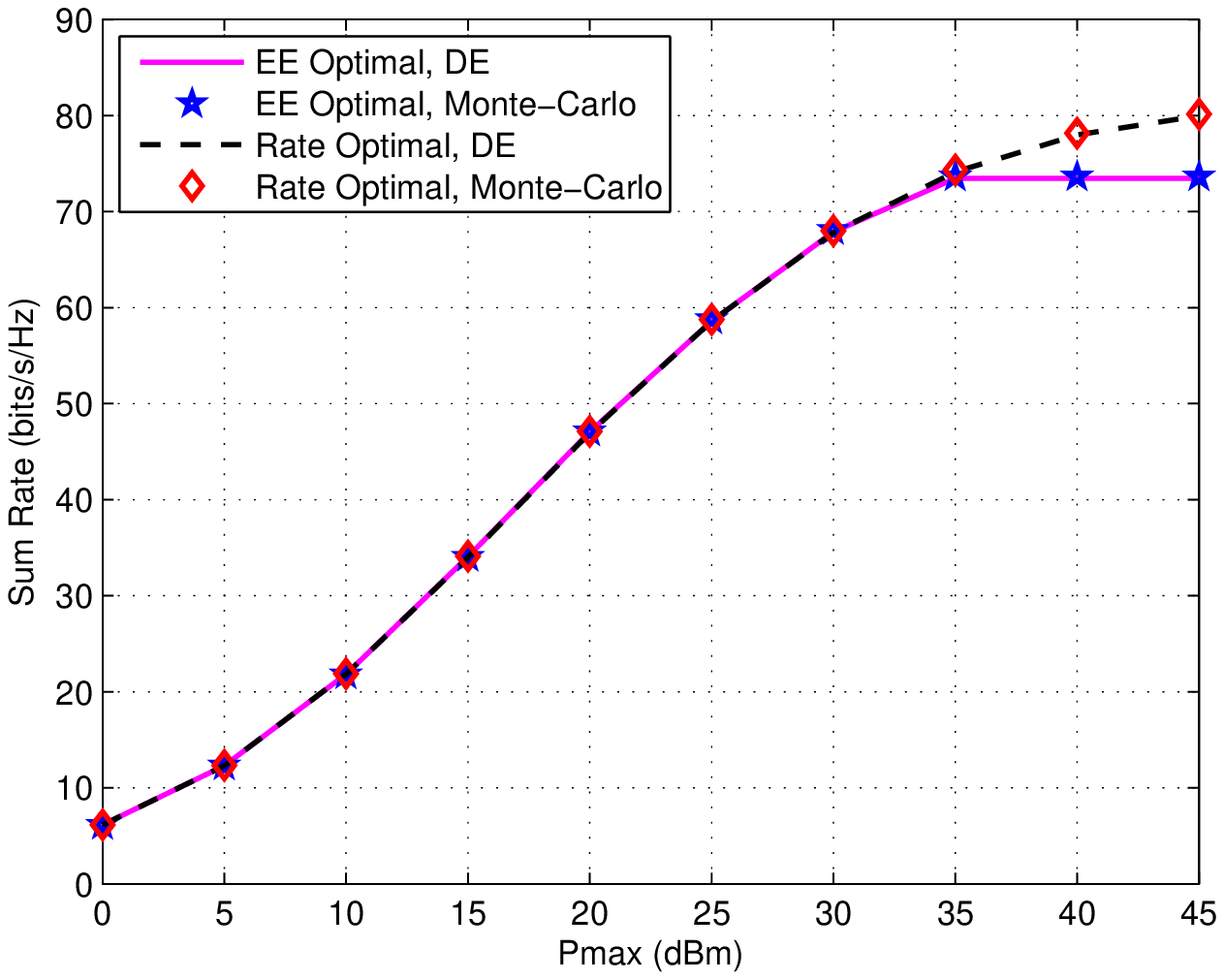}
\label{fig:Rate_performance}}
\caption{Comparison of EE performance and sum-rate performance versus $\Pmax$ with the aims of EE maximization and sum-rate maximization in \alref{alg:simplified}. (a) EE performance; (b) sum-rate performance.}
\label{fig:Comparison}
\end{figure*}

\begin{figure}
\centering
\includegraphics[width=\figsincolwid]{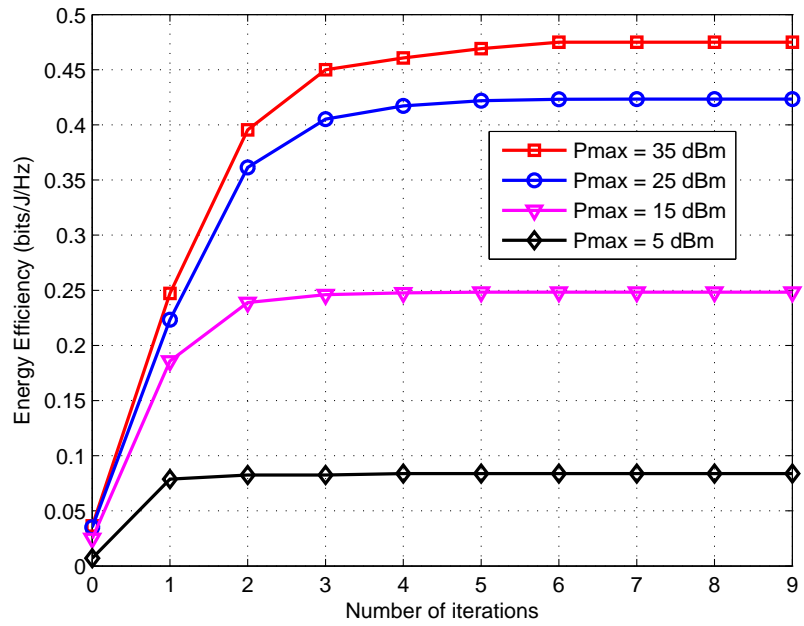}
\caption{Convergence behavior of \alref{alg:simplified} versus the numbers of iterations for different values of maximum power budget $\dnnot{P}{max}$.}
\label{fig:Convergence}
\end{figure}

\begin{figure}
\centering
\includegraphics[width=\figsincolwid]{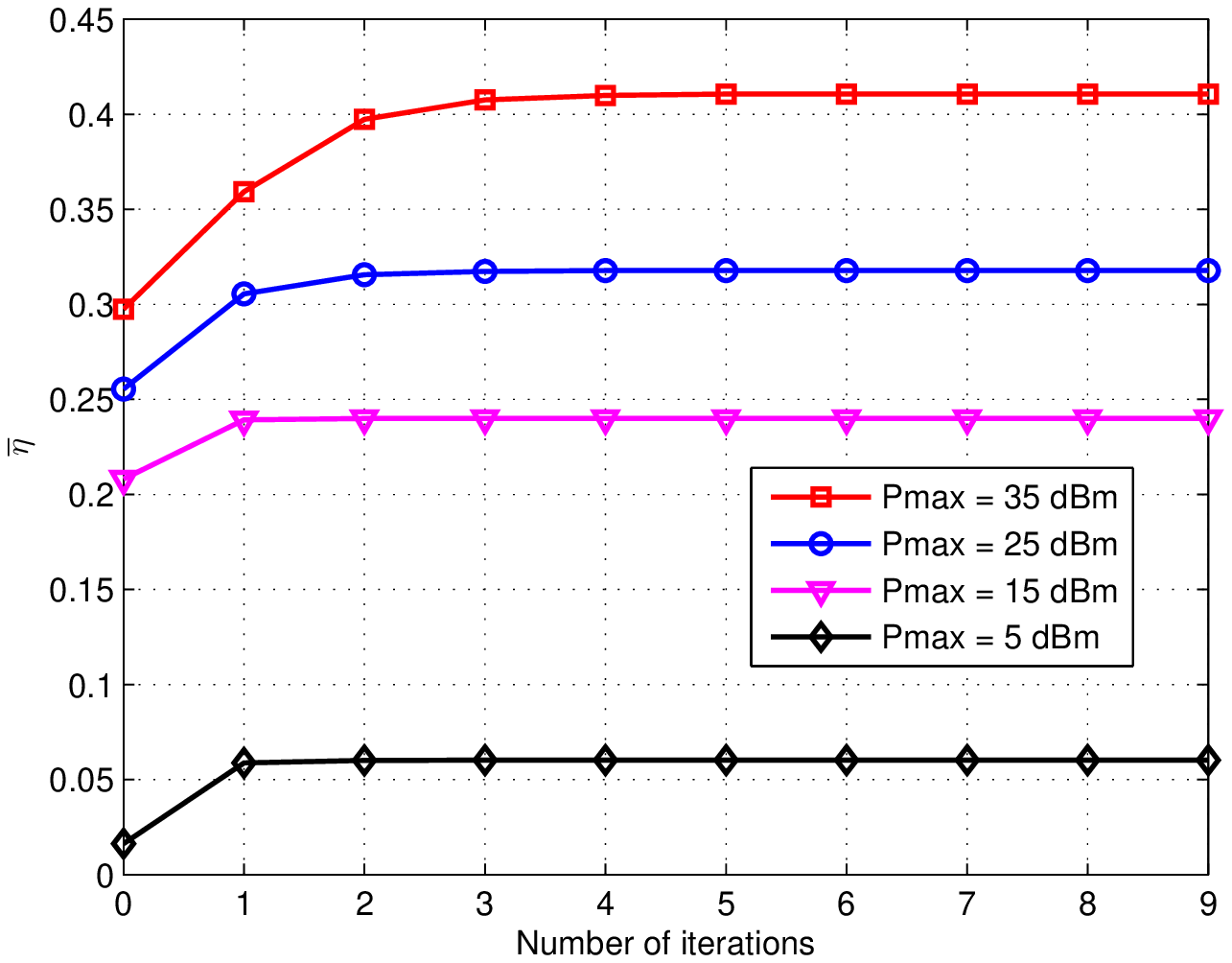}
\caption{Convergence behavior of \alref{alg:EE iterative water-filling} versus the numbers of iterations for different values of maximum power budget $\dnnot{P}{max}$.}
\label{fig:Convergence_alg3}
\end{figure}

The convergence behaviors of the iterative Algorithms \ref{alg:simplified} and \ref{alg:EE iterative water-filling} are presented in Figs. \ref{fig:Convergence} and \ref{fig:Convergence_alg3}, respectively. The results indicate that our proposed \alref{alg:simplified} has quick convergence performance and usually converges after only two or three iterations. In particular, the optimal performance can be approached after only one iteration for low $\Pmax$. We can also observe that the convergence rate of \alref{alg:simplified} becomes slightly slower when $\Pmax$ increases. From \figref{fig:Convergence_alg3}, we observe that \alref{alg:EE iterative water-filling} also converges fast in typical power budget regions.


\begin{figure}
\centering
\includegraphics[width=\figsincolwid]{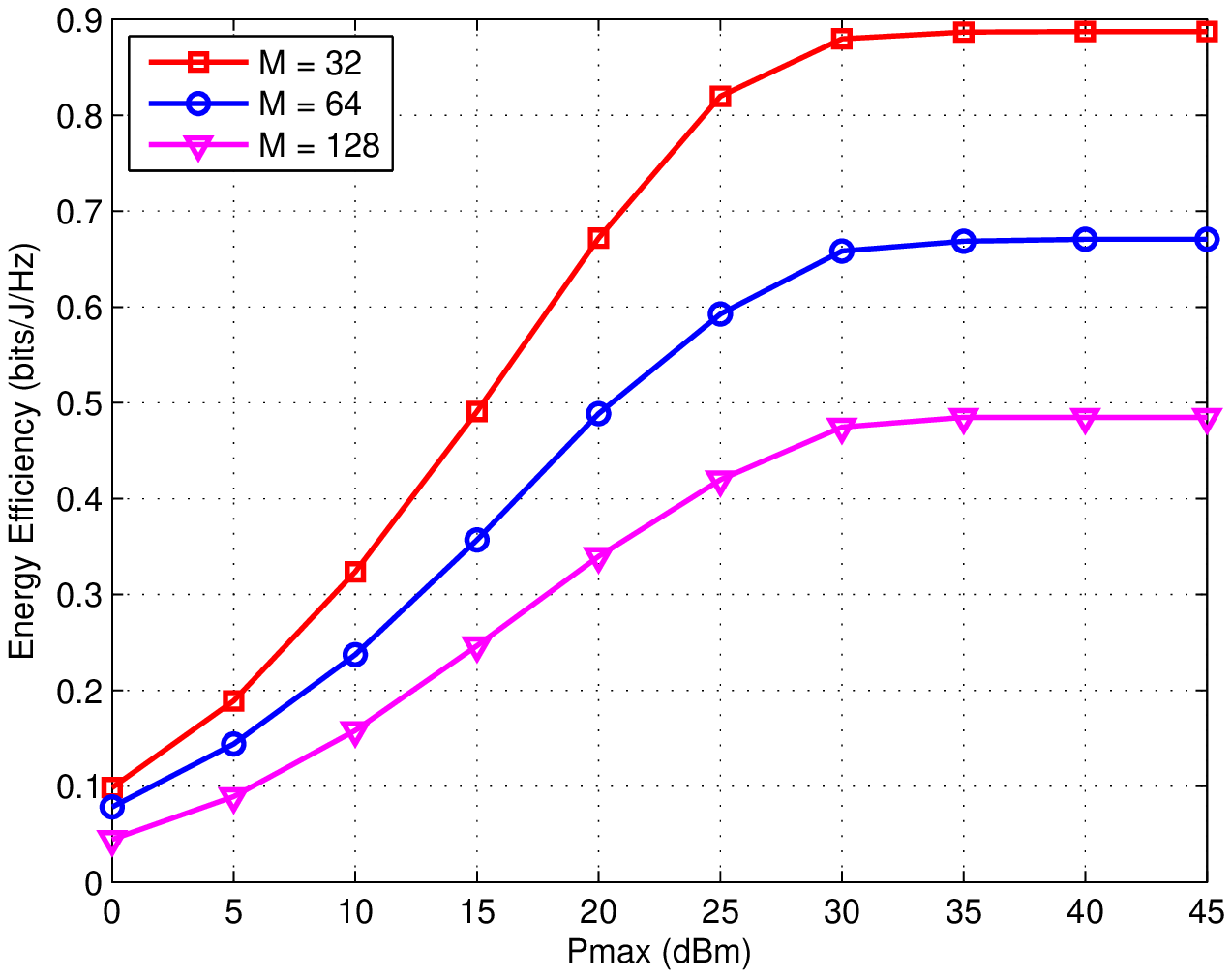}
\caption{The EE performance versus the maximum power budget $\Pmax$ for different numbers of BS antennas $M$.}
\label{fig:M_different}
\end{figure}

\figref{fig:M_different} depicts the EE performance of our proposed statistical CSI aided iterative \alref{alg:simplified} versus the number of BS antennas. We can observe a decreasing tendency of the EE value when the number of BS antennas increases, which is due to the linear increasing tendency in power consumption related to the number of BS antennas.

\begin{figure}
\centering
\includegraphics[width=\figsincolwid]{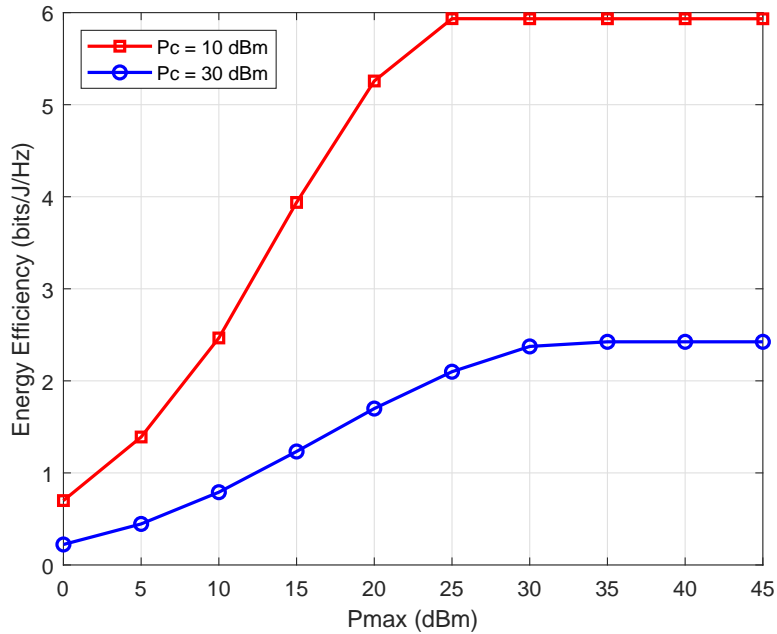}
\caption{The EE performance versus the maximum power budget $\Pmax$ for different values of circuit power consumption per antenna $\Pc$.}
\label{fig:Pc_different}
\end{figure}

\figref{fig:Pc_different} illustrates the EE performance of our proposed \alref{alg:simplified} for the two circuit power values per antenna of $\Pc$ = 10 dBm and $\Pc$ = 30 dBm, respectively. In order to clearly show the results for the two circuit power values in the same figure, we have magnified the EE value corresponding to the case of $\Pc$ = 30 dBm by a factor of $5$ as in \cite{zappone16energy}. We can observe that the EE performance improves as the circuit power per antenna decreases, which indicates that system EE will increase if $\Pc$ can be reduced. In addition, we can also observe that the EE saturation point is shifted to the right as $\Pc$ increases. This is due to the reason that for a higher $\Pc$, the transmit power will also increase before it becomes the dominant term in the denominator of the EE, and the optimal tradeoff between numerator and denominator is reached. Actually, if $\Pc \gg \Pmax$, the EE maximization approach will reduce to the sum-rate maximization approach.

\section{Conclusion}\label{sec:conclusion}
We have investigated single-cell massive MIMO downlink precoding under the EE maximization criterion with only statistical CSIT. We first showed the solution of the optimal transmit signal direction in a closed-form. Consequently, the maximum system EE for massive MIMO downlink could be acquired in the beam domain. Based on this conclusion, we reduced the complex transmit strategy design into a power allocation problem in the beam domain. Exploiting the MM algorithm and Dinkelbach's transform, a sequential algorithm was further proposed to solve such a power allocation problem, together with the reduction of computational complexity using the deterministic equivalent theory. Furthermore, we proposed a generalized iterative water-filling scheme via separating the constrained EE maximization problem into an unconstrained EE maximization problem and a constrained sum-rate maximization problem. We demonstrated by numerical results the EE improvement of our proposed EE optimization method over the sum-rate optimization method, especially in the high power budget regime.

\appendices

\section{Proof of \propref{theorem:beam_domain_optimal}}\label{app:A}

Denote by $\widetilde{\bQ}_k = \bV^H \bQ_k \bV(\forall k)$. Then, the sum rate of the system, which is also the numerator of the objective function in problem $\cF$, can be expressed as
\begin{align}\label{eq:rate_beam}
R_{\mathrm{sum}} = \sum\limits_{k = 1}^K R_k = \sum\limits_{k = 1}^K \left(\expect{ \log\det\left( \widetilde{\bK}_k + \bG_k \widetilde{\bQ}_k \bG_k^H\right)} -\log\det\left(\widetilde{\bK}_k\right) \right)
\end{align}
where
\begin{align}\label{eq:K_beam}
\widetilde{\bK}_k & = {\sigma}^2 \bI_{N_k} + \sum\limits_{i \neq k}^K {\expect{ \bG_k \widetilde{\bQ}_k \bG_k^H } } \ntb
& = {\sigma}^2 \bI_{N_k} + \sum\limits_{i \neq k}^K { \mathbf{\Pi}_k \left( \widetilde{\bQ}_i \right) }.
\end{align}
It is not difficult to check that the off-diagonal elements of $\bX$ do not affect the value of $\mathbf{\Pi}_k\left(\bX\right)$. Therefore, the element values of $\widetilde{\bK}_k$ are independent of the off-diagonal elements of $\widetilde{\bQ}_i(\forall i)$.

Following a similar line of reasoning of the proof in \cite[Theorem 1]{tulino2006capacity}, we define a diagonal matrix $\mathbf{D}_m \in \mathbb{R}^{M\times M}$ whose diagonal entries are $1$ except the $\left( m,m \right)$th entry which is $-1$. Then, the entries of $\mathbf{D}_m \widetilde{\bQ}_k \mathbf{D}_m$ are equivalent to those of $\widetilde{\bQ}_k$ except the off-diagonals in the $m$th row and $m$th column, whose signs are reversed. Thus, replacing $\widetilde{\bQ}_i$ with $\mathbf{D}_m \widetilde{\bQ}_i \mathbf{D}_m$ for $\forall i$ will not affect the value of $\widetilde{\bK}_k$ in \eqref{eq:K_beam}. Moreover, noticing that the zero-mean matrix $\bG_k$ is column-independent, its distribution will not be changed after being multiplied by a unitary matrix from either left or right. Since $\mathbf{D}_m$ is a unitary matrix, we can obtain that $\bG_k$ and $\bG_k \mathbf{D}_m$ have the same distribution, which yields
\begin{align}
& \qquad R_{\mathrm{sum}} \left( \widetilde{\bQ}_1, \ldots ,\widetilde{\bQ}_K \right)\ntb
& = \sum\limits_{k = 1}^K \left(\expect{ \log\det\left( \widetilde{\bK}_k +  \bG_k \widetilde{\bQ}_k \bG_k^H\right)} -\log\det\left(\widetilde{\bK}_k\right) \right)\ntb
& = \sum\limits_{k = 1}^K \left(\expect{ \log\det\left( \widetilde{\bK}_k +  \bG_k \mathbf{D}_m \widetilde{\bQ}_k \mathbf{D}_m \bG_k^H\right)} -\log\det\left(\widetilde{\bK}_k\right) \right)\ntb
& = R_{\mathrm{sum}}\left(\mathbf{D}_m \widetilde{\bQ}_1 \mathbf{D}_m , \ldots ,\mathbf{D}_m \widetilde{\bQ}_K \mathbf{D}_m \right).
\end{align}
Note that the matrix $\frac{1}{2}\left( \widetilde{\bQ}_k  + \mathbf{D}_m \widetilde{\bQ}_k \mathbf{D}_m\right)$ has entries equal to those of $\widetilde{\bQ}_k$ except for the off-diagonal entries in the $m$th row and $m$th column, which are all $0$. Moreover, invoking Jensen's inequality, we can have
\begin{align}
& \qquad R_{\mathrm{sum}} \left(\frac{1}{2}\left(\widetilde{\bQ}_1  + \mathbf{D}_m \widetilde{\bQ}_1 \mathbf{D}_m \right), \ldots ,\frac{1}{2}\left(\widetilde{\bQ}_K + \mathbf{D}_m \widetilde{\bQ}_K \mathbf{D}_m \right)\right)\ntb
& = \sum\limits_{k = 1}^K \left(\expect{ \log\det\left( \widetilde{\bK}_k +  \frac{1}{2}\bG_k\left(\widetilde{\bQ}_k + \mathbf{\                                                 D}_m \widetilde{\bQ}_k\mathbf{D}_m \right) \bG_k^H\right)} -\log\det\left(\widetilde{\bK}_k\right) \right)\ntb
& \ge \frac{1}{2}  \sum\limits_{k = 1}^K \left( \expect{ \log\det\left( \widetilde{\bK}_k + \bG_k \widetilde{\bQ}_k \bG_k^H\right)} -\log\det\left(\widetilde{\bK}_k\right) \right)\ntb
& \qquad +  \frac{1}{2} \sum\limits_{k = 1}^K \left( \expect{ \log\det\left( \widetilde{\bK}_k + \bG_k \mathbf{D}_m \widetilde{\bQ}_k \mathbf{D}_m \bG_k^H\right)} -\log\det\left(\widetilde{\bK}_k\right) \right)\ntb
& = \frac{1}{2}R_{\mathrm{sum}}\left(\widetilde{\bQ}_1 , \ldots ,\widetilde{\bQ}_K\right) + \frac{1}{2}R_{\mathrm{sum}}\left(\mathbf{D}_m \widetilde{\bQ}_1 \mathbf{D}_m , \ldots ,\mathbf{D}_m \widetilde{\bQ}_K \mathbf{D}_m \right)\ntb
& = R_{\mathrm{sum}} \left(\widetilde{\bQ}_1 , \ldots ,\widetilde{\bQ}_K \right).
\end{align}
The above inequality indicates that nulling the off-diagonal entries of any row and column of $\widetilde{\bQ}_1 , \ldots ,\widetilde{\bQ}_K $ will not decrease the sum rate in \eqref{eq:rate_beam}. Repeating this process for $m$ from $1$ to $M$, we can find that the numerator of problem $\cF$ is maximized when $\widetilde{\bQ}_1 , \ldots ,\widetilde{\bQ}_K $ are all diagonal. Meanwhile, changing the off-diagonal entries of any row and column of $\widetilde{\bQ}_1 , \ldots ,\widetilde{\bQ}_K $ does not change the denominator of the objective function in problem $\cF$. Therefore, the objective function in problem $\cF$ is maximized when $\widetilde{\bQ}_k= \bV^H \bQ_k \bV = \bV^H \mathbf{\Psi}_k \bLambda_k \mathbf{\Psi}_k^H \bV(\forall k)$ are all diagonal. This concludes the proof.

\section{Proof of \propref{prop:EE_solution}}\label{app:B}
The lagrangian function of problem \eqref{eq:unconstrained_problem} is defined as
\begin{align}\label{eq:lag1}
{\cal L} & =  { \sum\limits_{k=1}^{K}{\left(  {\Rbar}_{k}^{+}\left(\Lda\right) - R_{k}^{-}\left(\Lda^{(\ell)}\right) - {{\mathop{\rm tr}\nolimits} \left( {{{{\bf{\Delta }}^{(\ell)}_{k}}} \left( {{{\bf{\Lambda }}_k} - {{\bf{\Lambda }}^{(\ell)}_{k}}} \right)} \right)} \right)} } \ntb
& \qquad- {\overline \eta}^{(\ell)}_{(i)}\left({ \xi \sum\limits_{k=1}^{K}{\trr\left( \lambdak \right)} + M \Pc + \Ps }\right) + \sum\limits_{k = 1}^K {{\mathop{\rm tr}\nolimits} ({{\bf{\Psi }}_k}{{\bf{\Lambda }}_k})}
\end{align}
where the Lagrange multipliers ${{\bf{\Psi }}_k}\succeq \bzero$ depend on the problem constraints. The gradient of ${\Rbar}_{k}^{+}\left(\Lda\right)$ over ${\bf{\Lambda }}_{k}$ can be derived from \eqref{eq:DE} as
\begin{align}\label{eq:g_Rk}
\frac{\partial }{{\partial {{\bf{\Lambda }}_k}}}\Rbar_k^ + ({\bf{\Lambda }}) &= {\left( {{\bf{I}}_{M} + {{\bf{\Gamma }}_k}{{\bf{\Lambda }}_k}} \right)^{ - 1}}{{\bf{\Gamma }}_k} + \sum\limits_{m,n} {\frac{{\partial \Rbar_k^ + ({\bf{\Lambda }})}}{{\partial {{[{{\widetilde \eta }_k}({\bf{\Phi }}_k^{ - 1}{{\bf{\Lambda }}_k})]}_{m,n}}}}\frac{{\partial {{[{{\widetilde \eta }_k}({\bf{\Phi }}_k^{ - 1}{{\bf{\Lambda }}_k})]}_{m,n}}}}{{\partial {{\bf{\Lambda }}_k}}}} \ntb
 & \qquad + \sum\limits_{m,n} {\frac{{\partial \Rbar_k^ + ({\bf{\Lambda }})}}{{\partial {{[{\eta _k}({\bf{\widetilde \Phi }}_k^{ - 1}{\overline{\bf{K}}}_k^{ - 1})]}_{m,n}}}}\frac{{\partial {{[{\eta _k}({\bf{\widetilde \Phi }}_k^{ - 1}{\bf{\overline K}}_k^{ - 1})]}_{m,n}}}}{{\partial {{\bf{\Lambda }}_k}}}}.
\end{align}
Following an approach similar to proving Theorem 4 in \cite{Lu16Free}, we have
\begin{align}\label{eq:der2}
{\frac{{\partial \Rbar_k^ + ({\bf{\Lambda }})}}{{\partial {{[{{\widetilde \eta }_k}({\bf{\Phi }}_k^{ - 1}{{\bf{\Lambda }}_k})]}_{m,n}}}}} = 0\\
{\frac{{\partial \Rbar_k^ + ({\bf{\Lambda }})}}{{\partial {{[{\eta _k}({\bf{\widetilde \Phi }}_k^{ - 1}{\bf{\overline K}}_k^{ - 1})]}_{m,n}}}}}=0
\end{align}
which further leads to
\begin{align}\label{eq:g_k}
\frac{\partial }{{\partial {{\bf{\Lambda }}_k}}}\Rbar_{k}^{+} \left( \Lda \right)= {\left( {{\bf{I}}_{M} + {\bf{\Gamma }}_k{{\bf{\Lambda }}_k}} \right)^{ - 1}}{\bf{\Gamma }}_k.
\end{align}
In addition, the gradient of $\Rbar_{k}^{+}\left(\Lda\right)$ over $\Lda_{k'}(k' \ne k)$ is derived as
\begin{align}\label{eq:g_l}
\frac{\partial }{{\partial {{\bf{\Lambda }}_{k'}}}}\Rbar_{k}^{+}\left(\Lda\right) = \sum\limits_{n = 1}^{N_k} {\frac{{{{\hat {\bf{R}}}_{k,n}}}}{{\widetilde \gamma _{k,n} + {\sigma}^2 +  {\mathop{\rm tr}\nolimits} ({{\bf{\Lambda }}_{\backslash k}}{{\widehat {\bf{R}}}_{k,n}})}}}
\end{align}
where ${\widetilde \gamma _{k,n}}$ denotes the $n$th diagonal element of ${\bf{\widetilde \Gamma }}_k$. Then, from \eqref{eq:g_k} and \eqref{eq:g_l}, we have
\begin{align}
\frac{\partial }{{\partial {{\bf{\Lambda }}_a}}}\sum\limits_{k = 1}^K {{\overline R}_k^{+}\left(\Lda\right)}  = {\left( {{\bf{I}}_{M} + {\bf{\Gamma }}_a{{\bf{\Lambda }}_a}} \right)^{ - 1}}{\bf{\Gamma }}_a + \sum\limits_{k \ne a}^K {\sum\limits_{n = 1}^{N_k} {\frac{{{{\widehat {\bf{R}}}_{k,n}}}}{{\widetilde \gamma _{k,n} + {\sigma}^2 + {\mathop{\rm tr}\nolimits} ({{\bf{\Lambda }}_{\backslash k}}{{\widehat {\bf{R}}}_{k,n}})}}} }.
\end{align}
Due to the fact that $\Rbar_{k}^{+}\left(\Lda\right) $ is strictly concave with respect to $\Lda$, the KKT conditions of problem \eqref{eq:Dinkelbach_DE} are
\begin{align}\label{eq:KKT_1}
& \frac{{\partial {{\cal L}}}}{{\partial {\bf{\Lambda }}^{(\ell)}_{k,(i+1)}}} = \bzero,\quad k = 1, \ldots ,K \\ \label{eq:KKT_2}
& {\mathop{\rm tr}\nolimits} \left({\bf{\Psi }}^{(\ell)}_{k,(i+1)} {\bf{\Lambda }}^{(\ell)}_{k,(i+1)} \right) = 0,\quad {\bf{\Psi }}^{(\ell)}_{k,(i+1)} \succeq \bzero,\quad {\bf{\Lambda }}^{(\ell)}_{k,(i+1)} \succeq \bzero.
\end{align}
Note that problem in \eqref{eq:Dinkelbach_DE} is a convex program. Therefore, we can acquire its optimal solution ${{\bf{\Lambda }}^{(\ell)}_{k,(i+1)}}$ through solving the corresponding KKT conditions. From \eqref{eq:lag1} and \eqref{eq:der2}, we reformulate its first KKT condition in \eqref{eq:KKT_1} as
\begin{align}\label{eq:solution_matrix}
\frac{{\partial {{\cal L}}}}{{\partial {\bf{\Lambda }}^{(\ell)}_{k,(i+1)} }} & = {\left( {{\bf{I}}_{M} + {\bf{\Gamma }}^{(\ell)}_{k,(i+1)} {\bf{\Lambda }}^{(\ell)}_{k,(i+1)} } \right)^{ - 1}}{\bf{\Gamma }}^{(\ell)}_{k,(i+1)}- {{{\bf{\Delta }}^{(\ell)}_{k}}}  - { \xi {\overline \eta}_{(i)}^{(\ell)} } {\bf{I}}_{N_k} \ntb
&\qquad+ {\bf{\Psi }}^{(\ell)}_{k,(i+1)}+ \sum\limits_{k' \ne k}^K {\sum\limits_{n = 1}^{N_{k'}} {\frac{{{{\widehat {\bf R}}_{k',n}}}}{{\widetilde \gamma _{k',n,(i+1)}^{(\ell)} + {\sigma}^2 + {\mathop{\rm tr}\nolimits} ({\bf{\Lambda }}_{\backslash k',(i+1)}^{(\ell)}{{\widehat {\bf{R}} }_{k',n}})}}} } = \bf{0}.
\end{align}
Note that $\frac{{\partial {{\cal L}}}}{{\partial {\bf{\Lambda }}^{(\ell)}_{k,(i+1)}}}$ above is a diagonal matrix. Then, KKT condition \eqref{eq:solution_matrix} can be reduced to
\begin{align}\label{eq:KKT_app}
\left[ \frac{{\partial {{\cal L}}}}{{\partial {\Lda }^{(\ell)}_{k,(i+1)}}} \right]_{m,m} &  = {\frac{{\gamma _{k,m,(i+1)}^{(\ell)}}}{{1 + \gamma _{k,m,(i+1)}^{(\ell)}\lambda _{k,m,(i+1)}^{(\ell)}}}} - d^{(\ell)} _{k,m}  - { \xi {\overline \eta}^{(\ell)}_{(i)} } + \varphi^{(\ell)}_{k,(i+1)}\ntb
&\quad + \sum\limits_{k' \ne k}^K {\sum\limits_{n = 1}^{N_{k'}} {\frac{{{{\widehat r}_{k',m,n}}}}{{\widetilde \gamma _{k',n,(i+1)}^{(\ell)} + {\sigma}^2 +  {\mathop{\rm tr}\nolimits} ({\bf{\Lambda }}_{\backslash k',(i+1)}^{(\ell)}{{\widehat {\bf{R}} }_{k',n}})}}} } = 0,m = 1,2, \ldots ,M.
\end{align}
Therefore, we can observe that the KKT conditions in \eqref{eq:KKT_1} and \eqref{eq:KKT_2} are equal to those of the following problem
\begin{align}\label{eq:Dinkelbach_short_app}
{\bf{\Lambda }}^{(\ell)}_{(i + 1)} = \mathop {\arg \max }\limits_{{{\bf{\Lambda }}}}\quad& \sum\limits_{k = 1}^K {\Bigg( {\log \det \left( {{\bf{I}}_{M} + {{\bf{\Gamma }}_k}{{\bf{\Lambda }}_k}} \right)
+ \log \det \left( {{{{\bf{\widetilde \Gamma }}}_k} + {{{\bf{\overline K}}}_k\left(\Lda\right)}} \right)} } \ntb
&  \qquad{  - {\mathop{\rm tr}\nolimits} \left( {{{\bf{\Delta }}^{(\ell)}_{k}}{{\bf{\Lambda }}_k}} \right) - \xi {{\overline \eta}^{(\ell)}_{(i)}}{\mathop{\rm tr}\nolimits} \left( {{{\bf{\Lambda }}_k}} \right)} \Bigg)\ntb
{\mathrm{s.t.}} \quad
& \lambdak \succeq \bzero,\quad\forall k\in \K.
\end{align}
Note that \eqref{eq:Dinkelbach_short_app} is also a convex program, whose KKT conditions are equivalent to those of \eqref{eq:Dinkelbach_DE}. Solving the KKT conditions, we have
\begin{align}\label{eq:Solution}
\left\{ \begin{array}{l}
\frac{{\gamma _{k,m,(i+1)}^{(\ell)}}}{{1 + \gamma _{k,m,(i+1)}^{(\ell)}\lambda _{k,m,(i+1)}^{(\ell)}}} \\
\quad + \sum\limits_{k' \ne k}^K {\sum\limits_{n = 1}^{N_{k'}} {\frac{{{{\widehat r}_{k',m,n}}}}{{\widetilde \gamma _{k',n,(i+1)}^{(\ell)} + {\sigma}^2 + {\mathop{\rm tr}\nolimits} ({{\widehat {\bf{R}} }_{k',n}}{\bf{\Lambda }}_{\backslash k',(i+1)}^{(\ell)})}}} }= {d ^{(\ell)}_{k,m}} + \xi { {\overline \eta}^{(\ell)}_{(i)}},\quad\xi {{\overline \eta}^{(\ell)} _{(i)}} < \upsilon _{k,m,(i+1)}^{(\ell)} - {d^{(\ell)} _{k,m}} \\
\lambda _{k,m,(i+1)}^{(\ell)} = 0,\quad\quad\quad\quad\quad\quad\quad\quad\quad\quad\quad\quad\quad\quad\quad\quad\quad\quad\xi {{\overline \eta}^{(\ell)} _{(i)}} \ge \upsilon _{k,m,(i+1)}^{(\ell)} - d^{(\ell)} _{k,m}
\end{array} \right.
\end{align}
where the auxiliary variable $\upsilon _{k,m,(i+1)}^{(\ell)}$ is expressed as
\begin{align}
\upsilon _{k,m,(i+1)}^{(\ell)} = \gamma _{k,m,(i+1)}^{(\ell)} + \sum\limits_{k' \ne k}^K {\sum\limits_{n = 1}^{N_{k'}} {\frac{{{{\widehat r}_{k',m,n}}}}{{\widetilde \gamma _{k',n,(i+1)}^{(\ell)} + {\sigma}^2 + \sum\limits_{\scriptstyle(l',m')\hfill\atop
\scriptstyle \in {\cal{S}}(k,m,k')\hfill} {{{\widehat r}_{k',m',n}}\lambda _{l',m',(i+1)}^{(\ell)}} }}} }.
\end{align}
This concludes the proof.


\begin{thebibliography}{10}
\providecommand{\url}[1]{#1}
\csname url@samestyle\endcsname
\providecommand{\newblock}{\relax}
\providecommand{\bibinfo}[2]{#2}
\providecommand{\BIBentrySTDinterwordspacing}{\spaceskip=0pt\relax}
\providecommand{\BIBentryALTinterwordstretchfactor}{4}
\providecommand{\BIBentryALTinterwordspacing}{\spaceskip=\fontdimen2\font plus
\BIBentryALTinterwordstretchfactor\fontdimen3\font minus
  \fontdimen4\font\relax}
\providecommand{\BIBforeignlanguage}[2]{{%
\expandafter\ifx\csname l@#1\endcsname\relax
\typeout{** WARNING: IEEEtran.bst: No hyphenation pattern has been}%
\typeout{** loaded for the language `#1'. Using the pattern for}%
\typeout{** the default language instead.}%
\else
\language=\csname l@#1\endcsname
\fi
#2}}
\providecommand{\BIBdecl}{\relax}
\BIBdecl

\bibitem{Marzetta10Noncooperative}
T.~L. Marzetta, ``{Noncooperative cellular wireless with unlimited numbers of
  base station antennas},'' \emph{{IEEE} Trans. Wireless Commun.}, vol.~9,
  no.~11, pp. 3590--3600, Nov. 2010.

\bibitem{Sun2015Beam}
C.~Sun, X.~Q. Gao, S.~Jin, M.~Matthaiou, Z.~Ding, and C.~Xiao, ``Beam division
  multiple access transmission for massive {MIMO} communications,''
  \emph{{IEEE} Trans. Commun.}, vol.~63, no.~6, pp. 2170--2184, Jun. 2015.

\bibitem{marzetta2016fundamentals}
T.~L. Marzetta, \emph{Fundamentals of Massive {MIMO}}.\hskip 1em plus 0.5em
  minus 0.4em\relax New York, NY, USA: Cambridge Univ. Press, 2016.

\bibitem{bjornson2016massive}
E.~Bj{\"o}rnson, E.~G. Larsson, and T.~L. Marzetta, ``Massive {MIMO}: Ten myths
  and one critical question,'' \emph{{IEEE} Commun. Mag.}, vol.~54, no.~2, pp.
  114--123, Feb. 2016.

\bibitem{bjornson2017massive}
E.~Bj{\"o}rnson, J.~Hoydis, and L.~Sanguinetti, ``{Massive MIMO networks:
  Spectral, energy, and hardware efficiency},'' \emph{Found. Trends. Signal
  Process.}, vol.~11, no. 3-4, pp. 154--655, Nov. 2017.

\bibitem{zappone2015energy}
A.~Zappone and E.~Jorswieck, ``{Energy efficiency in wireless networks via
  fractional programming theory},'' \emph{{Found. Trends Commun. Inf. Theory}},
  vol.~11, no. 3-4, pp. 185--396, Jun. 2015.

\bibitem{bjornson2016deploying}
E.~Bj{\"o}rnson, L.~Sanguinetti, and M.~Kountouris, ``{Deploying dense networks
  for maximal energy efficiency: Small cells meet massive MIMO},'' \emph{{IEEE}
  J. Sel. Areas Commun.}, vol.~34, no.~4, pp. 832--847, Apr. 2016.

\bibitem{wang2016energy}
Y.~Wang, C.~Li, Y.~Huang, D.~Wang, T.~Ban, and L.~Yang, ``Energy-efficient
  optimization for downlink massive {MIMO} {FDD} systems with transmit-side
  channel correlation.'' \emph{{IEEE} Trans. Veh. Technol.}, vol.~65, no.~9,
  pp. 7228--7243, Sep. 2016.

\bibitem{Zappone2016energy}
A.~Zappone, L.~Sanguinetti, G.~Bacci, E.~Jorswieck, and M.~Debbah,
  ``Energy-efficient power control: A look at {5G} wireless technologies,''
  \emph{{IEEE} Trans. Signal Process.}, vol.~64, no.~7, pp. 1668--1683, Apr.
  2016.

\bibitem{pizzo2018network}
A.~Pizzo, D.~Verenzuela, L.~Sanguinetti, and E.~Bj{\"o}rnson, ``Network
  deployment for maximal energy efficiency in uplink with multislope path
  loss,'' \emph{{IEEE} Trans. Green Commun. Netw.}, vol.~2, no.~3, pp.
  735--750, Sep. 2018.

\bibitem{Vaezy2019energy}
H.~{Vaezy}, M.~J. {Omidi}, M.~M. {Naghsh}, and H.~{Yanikomeroglu}, ``Energy
  efficient transceiver design in {MIMO} interference channels: The selfish,
  unselfish, worst-case and robust methods,'' \emph{{IEEE} Trans. Commun.},
  vol.~67, no.~8, pp. 5377--5389, Aug. 2019.

\bibitem{xu2013energy}
J.~Xu and L.~Qiu, ``Energy efficiency optimization for {MIMO} broadcast
  channels,'' \emph{{IEEE} Trans. Wireless Commun.}, vol.~12, no.~2, pp.
  690--701, Feb. 2013.

\bibitem{he2014coordinated}
S.~He, Y.~Huang, L.~Yang, and B.~Ottersten, ``Coordinated multicell multiuser
  precoding for maximizing weighted sum energy efficiency,'' \emph{{IEEE}
  Trans. Signal Process.}, vol.~62, no.~3, pp. 741--751, Feb. 2014.

\bibitem{he2015energy}
S.~He, Y.~Huang, S.~Jin, and L.~Yang, ``Energy efficient coordinated
  beamforming design in multi-cell multicast networks,'' \emph{{IEEE} Commun.
  Lett.}, vol.~19, no.~6, pp. 985--988, Jun. 2015.

\bibitem{tervo2017energy}
O.~Tervo, A.~T{\"o}lli, M.~Juntti, and L.-N. Tran, ``Energy-efficient beam
  coordination strategies with rate-dependent processing power,'' \emph{{IEEE}
  Trans. Signal Process.}, vol.~65, no.~22, pp. 6097--6112, Nov. 2017.

\bibitem{tervo2018energy}
O.~Tervo, L.-N. Tran, H.~Pennanen, S.~Chatzinotas, B.~Ottersten, and M.~Juntti,
  ``Energy-efficient multicell multigroup multicasting with joint beamforming
  and antenna selection,'' \emph{{IEEE} Trans. Signal Process.}, vol.~66,
  no.~18, pp. 4904--4919, Sep. 2018.

\bibitem{choi2014downlink}
J.~Choi, D.~J. Love, and P.~Bidigare, ``Downlink training techniques for {FDD}
  massive {MIMO} systems: Open-loop and closed-loop training with memory,''
  \emph{{IEEE} J. Sel. Topics Signal Process.}, vol.~8, no.~5, pp. 802--814,
  Oct. 2014.

\bibitem{You15Pilot}
L.~You, X.~Q. Gao, X.-G. Xia, N.~Ma, and Y.~Peng, ``{Pilot reuse for massive
  MIMO transmission over spatially correlated Rayleigh fading channels},''
  \emph{{IEEE} Trans. Wireless Commun.}, vol.~14, no.~6, pp. 3352--3366, Jun.
  2015.

\bibitem{You16Channel}
L.~You, X.~Q. Gao, A.~L. Swindlehurst, and W.~Zhong, ``{Channel acquisition for
  massive MIMO-OFDM with adjustable phase shift pilots},'' \emph{{IEEE} Trans.
  Signal Process.}, vol.~64, no.~6, pp. 1461--1476, Mar. 2016.

\bibitem{wang2013precoder}
J.~Wang, M.~Matthaiou, S.~Jin, and X.~Q. Gao, ``{Precoder design for multiuser
  MISO systems exploiting statistical and outdated CSIT},'' \emph{{IEEE} Trans.
  Commun.}, vol.~61, no.~11, pp. 4551--4564, Nov. 2013.

\bibitem{wang2012statistical}
J.~Wang, S.~Jin, X.~Q. Gao, K.-K. Wong, and E.~Au, ``{Statistical
  eigenmode-based SDMA for two-user downlink},'' \emph{{IEEE} Trans. Signal
  Process.}, vol.~60, no.~10, pp. 5371--5383, Oct. 2012.

\bibitem{khalilsarai2019fdd}
M.~Barzegar~Khalilsarai, S.~Haghighatshoar, X.~Yi, and G.~Caire, ``{FDD massive
  MIMO via UL/DL channel covariance extrapolation and active channel
  sparsification},'' \emph{{IEEE} Trans. Wireless Commun.}, vol.~18, no.~1, pp.
  121--135, Jan. 2019.

\bibitem{xiong2019energy}
J.~Xiong, L.~You, X.~Yi, J.~Wang, W.~Wang, and X.~Q. Gao, ``Energy efficient
  precoding for massive {MIMO} downlink transmission with statistical {CSI},''
  in \emph{Proc. IEEE GLOBECOM}, Big Island, HI, USA, 2019, submitted.

\bibitem{Gao09Statistical}
X.~Q. Gao, B.~Jiang, X.~Li, A.~B. Gershman, and M.~R. McKay, ``{Statistical
  eigenmode transmission over jointly correlated MIMO channels},'' \emph{{IEEE}
  Trans. Inf. Theory}, vol.~55, no.~8, pp. 3735--3750, Aug. 2009.

\bibitem{You17BDMA}
L.~You, X.~Q. Gao, G.~Y. Li, X.-G. Xia, and N.~Ma, ``{BDMA for
  millimeter-wave/Terahertz massive MIMO transmission with per-beam
  synchronization},'' \emph{{IEEE} J. Sel. Areas Commun.}, vol.~35, no.~7, pp.
  1550--1563, Jul. 2017.

\bibitem{Adhikary13Joint}
A.~Adhikary, J.~Nam, J.-Y. Ahn, and G.~Caire, ``{Joint spatial division and
  multiplexing---The large-scale array regime},'' \emph{{IEEE} Trans. Inf.
  Theory}, vol.~59, no.~10, pp. 6441--6463, Oct. 2013.

\bibitem{Hassibi2003How}
B.~Hassibi and B.~M. Hochwald, ``How much training is needed in
  multiple-antenna wireless links?'' \emph{{IEEE} Trans. Inf. Theory}, vol.~49,
  no.~4, pp. 951--963, Apr. 2003.

\bibitem{Lu2019Robust}
A.-A. {Lu}, X.~Q. {Gao}, W.~{Zhong}, C.~{Xiao}, and X.~{Meng}, ``{Robust
  transmission for massive MIMO downlink with imperfect CSI},'' \emph{{IEEE}
  Trans. Commun.}, vol.~67, no.~8, pp. 5362--5376, Aug. 2019.

\bibitem{wu18beam}
W.~Wu, X.~Q. Gao, Y.~Wu, and C.~Xiao, ``Beam domain secure transmission for
  massive {MIMO} communications,'' \emph{{IEEE} Trans. Veh. Technol.}, vol.~67,
  no.~8, pp. 7113--7127, Aug. 2018.

\bibitem{Marks1978A}
B.~R. Marks and G.~P. Wright, ``A general inner approximation algorithm for
  nonconvex mathematical programs,'' \emph{Oper. Res.}, vol.~26, no.~4, pp.
  681--683, Aug. 1978.

\bibitem{Beck2010A}
A.~Beck, A.~Ben-Tal, and L.~Tetruashvili, ``A sequential parametric convex
  approximation method with applications to nonconvex truss topology design
  problems,'' \emph{J. Glob. Optim.}, vol.~47, no.~1, pp. 29--51, May 2010.

\bibitem{Razaviyayn2012A}
M.~Razaviyayn, M.~Hong, and Z.-Q. Luo, ``A unified convergence analysis of
  block successive minimization methods for nonsmooth optimization,''
  \emph{{SIAM} J. Optim.}, vol.~23, no.~2, pp. 1126--1153, Jun. 2013.

\bibitem{sun2017majorization}
Y.~Sun, P.~Babu, and D.~P. Palomar, ``Majorization-minimization algorithms in
  signal processing, communications, and machine learning,'' \emph{{IEEE}
  Trans. Signal Process.}, vol.~65, no.~3, pp. 794--816, Feb. 2017.

\bibitem{shen2018fractional}
K.~Shen and W.~Yu, ``{Fractional programming for communication systems---Part
  I: Power control and beamforming},'' \emph{{IEEE} Trans. Signal Process.},
  vol.~66, no.~10, pp. 2616--2630, May 2018.

\bibitem{Boyd04Convex}
S.~Boyd and L.~Vandenberghe, \emph{{Convex Optimization}}.\hskip 1em plus 0.5em
  minus 0.4em\relax New York, NY, USA: Cambridge Univ. Press, 2004.

\bibitem{Couillet11Random}
R.~Couillet and M.~Debbah, \emph{{Random Matrix Methods for Wireless
  Communications}}.\hskip 1em plus 0.5em minus 0.4em\relax New York, NY, USA:
  Cambridge Univ. Press, 2011.

\bibitem{Lu16Free}
A.-A. Lu, X.~Q. Gao, and C.~Xiao, ``{Free deterministic equivalents for the
  analysis of MIMO multiple access channel},'' \emph{{IEEE} Trans. Inf.
  Theory}, vol.~62, no.~8, pp. 4604--4629, Aug. 2016.

\bibitem{Dumont2010On}
J.~Dumont, S.~Lasaulce, S.~Lasaulce, P.~Loubaton, and J.~Najim, ``On the
  capacity achieving covariance matrix for {Rician MIMO} channels: {An}
  asymptotic approach,'' \emph{{IEEE} Trans. Inf. Theory}, vol.~56, no.~3, pp.
  1048--1069, Jul. 2010.

\bibitem{Dupuy2011On}
F.~Dupuy and P.~Loubaton, ``On the capacity achieving covariance matrix for
  frequency selective {MIMO} channels using the asymptotic approach,''
  \emph{{IEEE} Trans. Inf. Theory}, vol.~57, no.~9, pp. 5737--5753, Sep. 2011.

\bibitem{chong2011analytical}
Z.~Chong and E.~Jorswieck, ``Analytical foundation for energy efficiency
  optimisation in cellular networks with elastic traffic,'' in \emph{Proc.
  MOBILIGHT.}, Bilbao, Spain, 2011, pp. 18--29.

\bibitem{cormen2009introduction}
T.~H. Cormen, C.~E. Leiserson, R.~L. Rivest, and C.~Stein, \emph{{Introduction
  to Algorithms}}, 3rd~ed.\hskip 1em plus 0.5em minus 0.4em\relax Cambridge,
  MA, USA: MIT press, 2009.

\bibitem{sun2017bdma}
C.~Sun, X.~Q. Gao, and Z.~Ding, ``{BDMA} in multicell massive {MIMO}
  communications: Power allocation algorithms,'' \emph{{IEEE} Trans. Signal
  Process.}, vol.~65, no.~11, pp. 2962--2974, Jun. 2017.

\bibitem{jaeckel2014quadriga}
S.~Jaeckel, L.~Raschkowski, K.~B{\"o}rner, and L.~Thiele, ``{QuaDRiGa}: A {3-D}
  multi-cell channel model with time evolution for enabling virtual field
  trials,'' \emph{{IEEE} Trans. Antennas Propag.}, vol.~62, no.~6, pp.
  3242--3256, Jun. 2014.

\bibitem{he2013coordinated}
S.~He, Y.~Huang, S.~Jin, and L.~Yang, ``Coordinated beamforming for energy
  efficient transmission in multicell multiuser systems,'' \emph{{IEEE} Trans.
  Commun.}, vol.~61, no.~12, pp. 4961--4971, Dec. 2013.

\bibitem{zappone16energy}
A.~Zappone, P.-H. Lin, and E.~Jorswieck, ``Energy efficiency of confidential
  multi-antenna systems with artificial noise and statistical {CSI},''
  \emph{{IEEE} J. Sel. Topics Signal Process.}, vol.~10, no.~8, pp. 1462--1477,
  Dec. 2016.

\bibitem{tulino2006capacity}
A.~M. Tulino, A.~Lozano, and S.~Verd{\'u}, ``Capacity-achieving input
  covariance for single-user multi-antenna channels,'' \emph{{IEEE} Trans.
  Wireless Commun.}, vol.~5, no.~3, pp. 662--671, Apr. 2006.

\end{thebibliography}


\end{document}